\documentclass[%
 twocolumn,
 table,
 superscriptaddress,
 amsmath,
 amssymb,
 aps,
 pra
]{revtex4-2}

\usepackage[utf8]{inputenc}
\usepackage[T1]{fontenc}
\usepackage{amssymb}
\usepackage{mathtools}
\usepackage{bbold}
\usepackage{bm}
\usepackage[position=top, caption=false]{subfig}
\usepackage[table,x11names,dvipsnames,table]{xcolor}
\usepackage{tikz}
\usetikzlibrary {shapes.geometric}

\usepackage{amsthm}
\usepackage{amsmath}
\usepackage{mathtools}
\usepackage{mathrsfs}
\usepackage{physics}
\usepackage{nicefrac}
\usepackage{xfrac}
\usepackage{graphicx}
\usepackage{csvsimple}
\usepackage{longtable}

\usepackage{algorithm}
\usepackage{algpseudocode}
\usepackage[left=15mm,right=15mm,top=20mm,columnsep=15pt]{geometry}
\usepackage[pdftex, pdftitle={Article}, pdfauthor={Author}]{hyperref}
\usepackage[noabbrev]{cleveref}
\creflabelformat{equation}{#2\textup{#1}#3}
\theoremstyle{definition}

\newtheorem{innercustomthm}{Result}
\newenvironment{customthm}[1]
  {\renewcommand\theinnercustomthm{#1}\innercustomthm}
  {\endinnercustomthm}

\begin{document}

\title{Teleporting two-qubit entanglement across 19 qubits on a superconducting quantum computer}



\author{Haiyue Kang}
\email{haiyuek@student.unimelb.edu.au}
\affiliation{School of Physics, The University of Melbourne, Parkville, Victoria 3010, Australia}
\author{John F. Kam}
\email{john.kam@monash.edu}
\affiliation{School of Physics \& Astronomy, Monash University, Clayton, Victoria 3800, Australia}
\author{Gary J. Mooney}
\email{mooney.g@unimelb.edu.au}
\affiliation{School of Physics, The University of Melbourne, Parkville, Victoria 3010, Australia}
\author{Lloyd C. L. Hollenberg}
\email{lloydch@unimelb.edu.au}
\affiliation{School of Physics, The University of Melbourne, Parkville, Victoria 3010, Australia}
\date{\today}

\begin{abstract} 
Quantum teleportation is not merely a fascinating corollary of quantum entanglement, it also finds utility in quantum processing and circuit compilation. In this paper, we measure and track
the entanglement and fidelity of two-qubit states prepared on a 127-qubit IBM Quantum computer, as one of the qubits is teleported across 19 qubits. We design, evaluate and compare two distinct approaches to teleportation: post-selected measurement categorisation and dynamic circuit corrections based on mid-circuit measurements, and compare with direct state transportation using SWAP gates. 
By optimally choosing the teleportation path which exhibits the highest total negativity entanglement measure across nearest-neighbour pairs, we show the entanglement of a two-qubit graph state is sustained after at least 19 hops in teleportation using the post-selection approach and 17 hops using the dynamic circuit approach. We observe a higher level of teleported entanglement in paths determined from two-qubit negativities compared to those obtained from gate errors, demonstrating an advantage in using the negativity map over the gate error map for compiling quantum circuits.

\end{abstract}

\maketitle

\section{Introduction} \label{sec:introduction}
To fully unleash the potential of quantum algorithms that cannot be time-efficiently executed classically and have applications in many fields \cite{shor, grover, VQE, QAOA, QML, QML_2, quantum_computational_chemistry, quantum_biology, quantum_finance}, the generation of large-scale entanglement on real quantum devices is an essential prerequisite. Hence, to demonstrate the utility of quantum computers, the generation and verification of large-scale quantum entanglement is often seen as a key metric \cite{classical_simulation_1, classical_simulation_2, classical_simulation_3}. In the current era of Noisy Intermediate-Scale Quantum (NISQ) devices, researchers have produced ever larger entangled states and improved entanglement measurement strategies, including the demonstration of bipartite entanglement \cite{bipartite_20, bipartite_65}, Genuine Multipartite Entanglement (GME) on Greenberger-Horne-Zelinger (GHZ) and cluster states up to 32 \cite{GHZ_32, fidel_paper} and 51 \cite{cao51entanglement} qubits, respectively.

In this work, we go beyond the generation and verification of entanglement and consider the utility of entanglement via the realisation of quantum state teleportation across multiple qubits. Such utilisation of entanglement is an important ingredient in hardware circuit compilation for both qubit pathfinding and implementing long-range gates, which are important elements in creating large-scale entanglement and benchmarking the capabilities of a quantum device. Since Bennett et al.\cite{Teleportation} first proposed quantum state teleportation, efforts have been made to physically realise and demonstrate quantum teleportation since two decades ago. The phenomenon of quantum teleportation was first experimentally verified by \cite{first_experimental_teleportation} in 1997, with teleportation of a polarised state in \cite{teleportation_of_polarization_state}. In addition, potential applications that use the principle of quantum teleportation have also been discovered, such as dense coding that allows one to communicate two bits of classical information by just sending one bit of classical information \cite{dense_coding}, quantum encrypted communication \cite{quantum_communication}, and detection of entanglement by teleporting the quantum state through photon modes of an extended Eienstein-Podolsky-Rosen (EPR) state \cite{detect_entanglement_via_teleportation}. As time progressed, researchers have successfully performed teleportation onto distant qubits \cite{teleportation_on_distant_qubits, teleportation_on_distant_qubits2}. Lately, there have been breakthroughs in long-range quantum gates based on dynamic measurements that teleport locally connected gates \cite{long_range_gates_2, long_range_gates} that also share a similar principle as entanglement swapping \cite{entanglement_swapping}. However, there still lacks sufficient 
verification of teleportation on entangled states at a level of a two-digit number of qubit hops.

\begin{figure*}[hbtp]
     \includegraphics[width=1\linewidth]{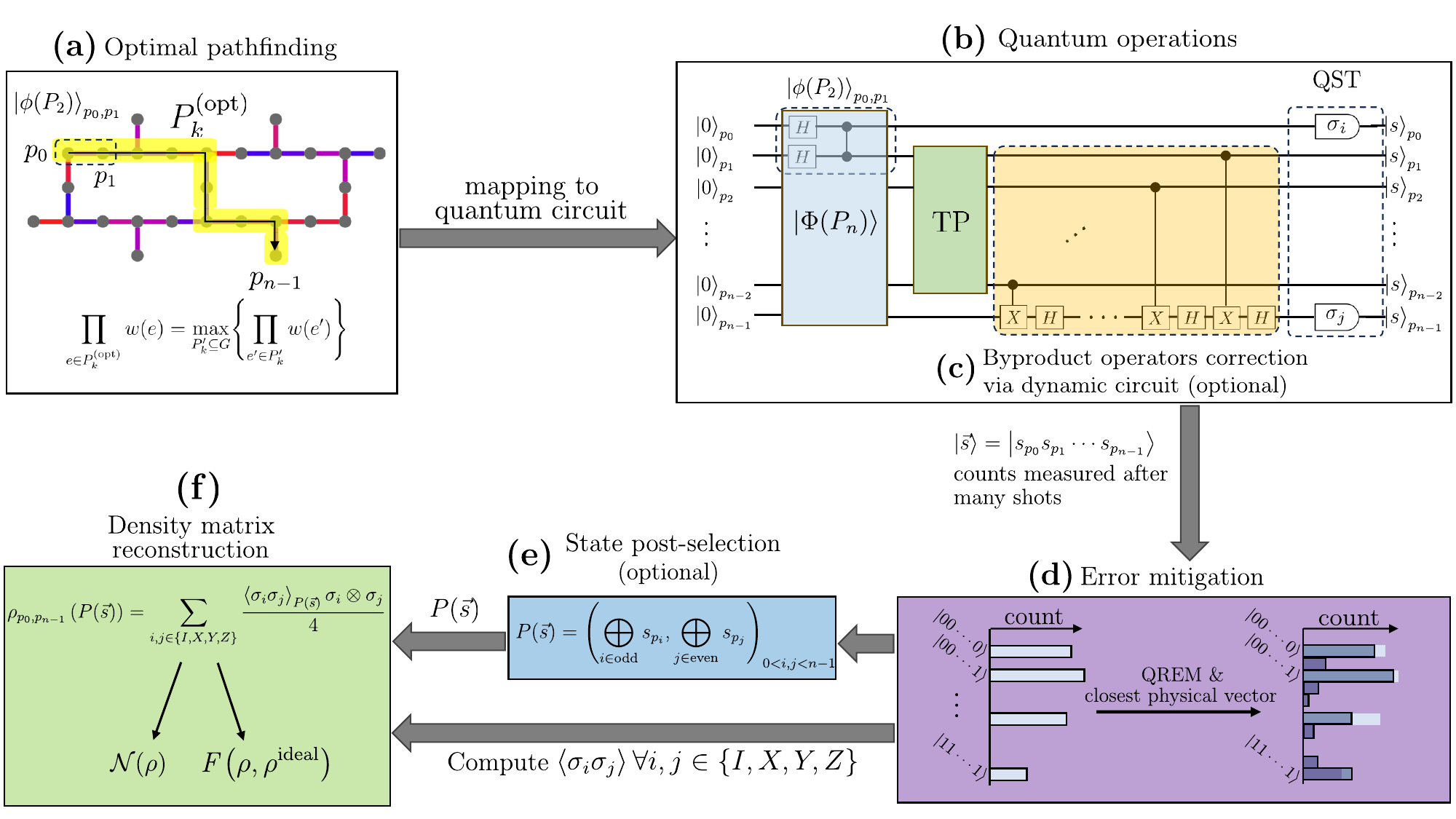}
     \caption{Overview of the teleportation workflow. \textbf{(a)} Optimal pathfinding: as in \Cref{sec: Teleportation path optimisation}, the line of qubits on the quantum device of a given path length $k$ is calculated based on maximising the edge weight product. \textbf{(b)} Quantum operations: a one-dimensional graph state is generated on the quantum circuit and the teleportation (TP) of the component of $\ket{\phi(P_2)}$ on $p_1$ to $p_{n-1}$ is performed through the measurement of intermediate qubits in the Pauli-$X$ basis as described in \Cref{sec: Teleportation}. At the end, the teleported state is measured in every Pauli basis combination to perform Quantum State Tomography (QST). \textbf{(c)} Byproduct operators correction applied via dynamic circuit described in \Cref{sec: dynamic circuits} as a mode of teleportation can be incorporated into the circuit to undo the local transformations acquired in the protocol. \textbf{(d)} Error Mitigation: applies Quantum Readout Error Mitigation (QREM) and removes unphysical negative counts that can arise after mitigation, as described in \Cref{sec: Entanglement Measurements}. \textbf{(e)} State post-selection described in \Cref{sec: post-selected categorisation} as another mode of teleportation used to discriminate the different projected Bell pairs. \textbf{(f)} Density matrix reconstruction of the teleported two-qubit state to calculate their corresponding negativity or fidelity.
     \label{fig:working flow chart}}
\end{figure*}
Here, we implement a teleportation protocol \cite{dynamic_circuit} based on single-qubit measurements on entangled graph states. The 
demonstrations were conducted on the IBM Quantum device \textit{ibm\_sherbrooke} consisting of 127 superconducting transmon qubits \cite{transmon_qubits} using the workflow shown in \Cref{fig:working flow chart}. We characterise and sustain the entanglement of two-qubit graph states that have one of their qubits teleported along the one-dimensional qubit paths up to 19 hops. It is found that a higher level of entanglement can be sustained if the paths used for teleportation are selected based on optimised net negativity, further highlighting the importance of entanglement benchmarking on a quantum device using bipartite negativities \cite{fidel_paper}. We implement two approaches to teleportation: post-selection and dynamic circuits. Post-selection preserves the form of the teleported two-qubit state by categorising measurements, while dynamic circuits correct the teleported two-qubit state to its original form using mid-circuit measurements. By comparing the robustness of these strategies with direct state transportation via a sequence of SWAP gates, we find that the post-selection method outperforms the others in terms of the preservation of entanglement.

\section{Teleportation along one-dimensional graph states}\label{sec: Teleportation}
\begin{figure}[t]
     \centering
     \includegraphics[width=1\linewidth]{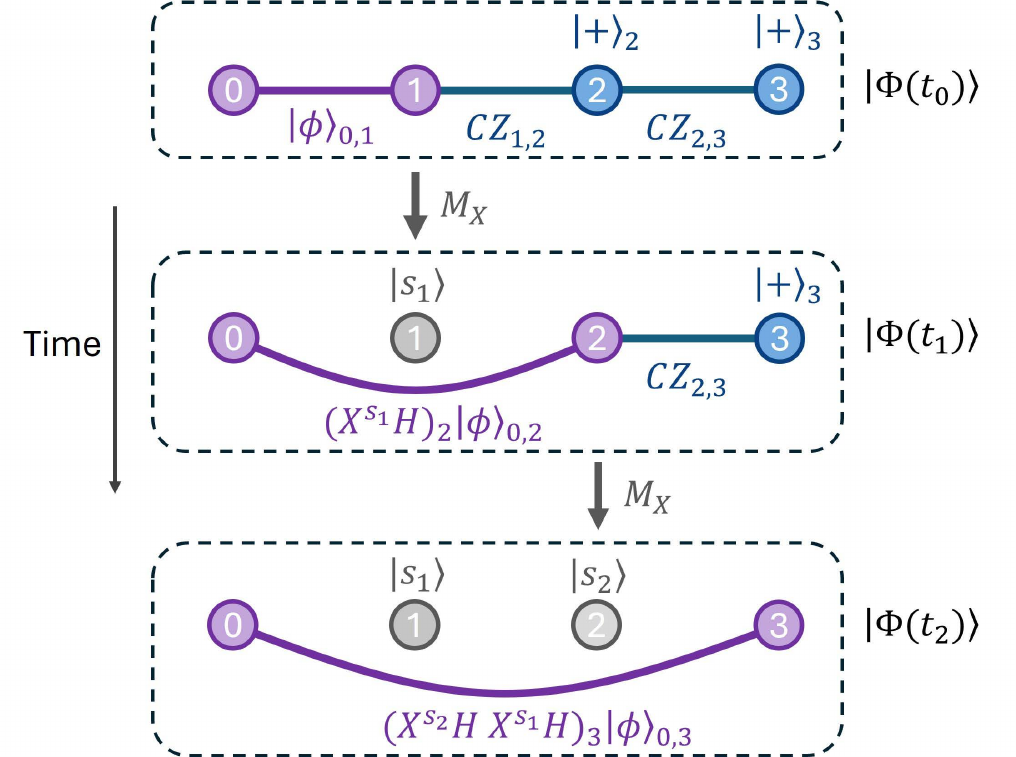}
     \caption[]{Demonstration of steps to teleport the component of $\ket{\phi}_{0,1}$ on qubit 1, which is entangled with qubit 0--illustrated in purple--towards the rightmost qubit. First at $t_{0}$, initialise the two-qubit state $\ket{\phi}_{0,1}$ and entangle with Pauli-X basis eigenstates $\ket{+}$ via $CZ$ gates, producing an overall state $\ket{\Phi(t_{0})}=CZ_{2,3}CZ_{1,2}\ket{\phi}_{0,1}\ket{+}_{2}\ket{+}_{3}$; Measuring qubit 1 in the Pauli-X basis where $M_{X_{1}}=M_{1,s_{1}}H_{1}$ (see \Cref{eq: measurement operator} in \Cref{appendix} for details), yielding a measurement result of $\ket{s_{1}}$, where $s_{1}=0$ or 1, producing the state $\ket{\Phi(t_{1})}=\ket{s_{1}}\otimes CZ_{2,3}(X^{s_{1}}H)_{2}\ket{\phi}_{0,2}\ket{+}_{3}$; Finally, another measurement on qubit 2 yields result $s_{2}$, leaving the teleported state at $t_{2}$ to be the same state as $\ket{\phi}$ up to some local transformations, namely $\ket{\Phi(t_{2})}=\ket{s_{1}s_{2}}\otimes(X^{s_{2}}HX^{s_{1}}H)_{3}\ket{\phi}_{0,3}$. Since the local transformations are known after measurements, they can be used to retrieve the original state $\ket{\phi}_{0,3}$ now defined over qubits 0 and 3. This procedure can be generalised to teleport the qubit 1 of the original entangled state across a path with an arbitrary number of nodes connected by $CZ$ gates.}
     \label{fig:teleportation}
\end{figure}
To prepare for quantum teleportation, one of the necessary prerequisites is the establishment of an entangled state between the teleporter and receiver sites. We generate a pair of entangled qubits and a path of entangled qubits with the intention of teleporting one of the qubits from the pair along the path. The entangled path is prepared as a graph state that comprises the second qubit, subsequent conduit `hopping' qubits, and the final qubit at the end receiver site. In the following, we trace through the theoretical framework of our setup by initially considering the teleportation of a qubit from an arbitrary two-qubit state. As shown in \Cref{fig:teleportation}, an arbitrary quantum state $\ket{\phi}_{0,1}=U(\ket{\alpha}\otimes\ket{\beta})$ is prepared on qubit 0 and 1, and teleported along a path of qubits that are already entangled via Controlled-$Z$ ($CZ$) gates acting on $\ket{+}$ states. By measuring qubit 1 of the path in the Pauli-$X$ basis, the information on that qubit is teleported to its neighbouring qubit 2 prepared in $\ket{+}$, while acquiring a local transformation depending on the measurement outcome applied to the state. The process can be repeated by measuring qubit 2 and treating the next-neighbouring qubit as the receiver. More importantly, while qubit 1 is being teleported to a target qubit, any form of entanglement previously established with other qubits is preserved. This feature allows us to perform entanglement characterisation over distant qubits by teleporting the state initially prepared on adjacent qubits.

We describe the natural evolution of the quantum state for the measurement-based teleportation protocol as follows (see \Cref{appendix} for details). In the Appendix, we derive the following results: Let $\ket{\phi}_{0,1}$ be an arbitrary two-qubit state prepared on qubits 0 and 1 that is entangled with--using a $CZ$ gate--a path graph state of $n-2$ qubits to produce a state with the form: 
\begin{equation}
    \ket{\Phi}=\left(\prod\limits_{i=1}^{n-2}{CZ_{i,i+1}}\right)\bigotimes\limits_{j=2}^{n-1}\ket{+}_{j}\otimes\ket{\phi}_{0,1}
\end{equation}
By measuring all intermediate qubits from label $1$ to $n-2$ in the $X$-basis with outcome $\vec{s}=(s_{1},s_{2},\cdots,s_{n-2})^{T}$, the qubit 1 component of state $\ket{\phi}_{0,1}$ is teleported along the path to qubit $n-1$ to produce $\ket{\phi
}_{0,n-1}$, while acquiring some local transformations depending on the measured results 
$\vec{s}$, which arrives at $\ket{\Phi}^{\text{(Teleported)}}=\left(\prod\limits_{i=1}^{n-2}X_{n-1}^{s_{i}}H_{n-1}\right)\ket{\phi}_{0,n-1}\otimes\ket{\vec{s}}$ that agrees with the result in \cite{dynamic_circuit}.

Next, we begin the description from the general definition of a graph state. According to the construction, a graph state mapped from a graph $G=(V,E)$ with vertices set $V$ and edge set $E$ corresponding to two-qubit connections, is defined as
\begin{equation}
    \ket{G}\coloneqq \prod\limits_{(i,j)\in E}{CZ_{i,j}{\ket{+}}^{\otimes n}},
\end{equation}
where $n=\abs{V}$ is the number of qubits, and $CZ_{i,j}$ is the controlled-$Z$ gate acting on edge $(i,j)$ of the graph $G$. We can set up the two-qubit state $\ket{\phi}_{0,1}$ to be teleported as a simple graph state on path $P_{2}$ with two vertices and an edge connecting them,
\begin{equation}
    \ket{\phi}_{0,1}=\frac{1}{2}(\ket{00}+\ket{01}+\ket{10}-\ket{11})_{0,1}\coloneqq\ket{\phi(P_{2})}_{0,1}.
\end{equation}
The prepared $n$-qubit state (where $n>2$) for the teleportation procedure $\ket{\Phi}$ is thus $\ket{\phi}_{0,1}$ entangled with an end qubit of a path graph state of $n-2$ qubits via a $CZ$ gate, resulting in a path graph state defined on the path graph $P_n$. The explicit form of this path graph state on $P_n$ with vertices $V(P_{n})=\{0,1\cdots,n-1\}$, and edges $E(P_{n})=\{(0,1),(1,2),\cdots(n-2,n-1)\}$ can be obtained as
\begin{equation}
    \ket{\Phi(P_{n})}=\frac{1}{\sqrt{2^{n}}}\sum\limits_{\vec{x}\in \{0,1\}^{\otimes n}}{\prod\limits_{i=0}^{n-2}{e^{i\pi x_{i}x_{i+1}}\ket{\vec{x}}}}.
\end{equation}
where $\ket{\vec{x}}=\ket{x_{0},x_{1},\cdots, x_{n-1}}$, and $x_{i}$ is the $i^{\text{th}}$ binary value in $\vec{x}$. The product of $e^{i\pi x_{i}x_{i+1}}$ is the phase picked up from every component $\ket{\vec{x}}$ after applying $CZ$ gate to all nearest-neighbour qubits.

From Result \ref{thm:teleportation across multiple qubit}, the teleported two-qubit graph state acquires local transformations in the following form
\vspace{10pt}
\begin{equation}\label{eq: local transformations_1}
    \ket{\psi}^{(\vec{s},n)}_{0,n-1}\coloneqq(X^{s_{n-2}}H\cdots X^{s_{2}}HX^{s_{1}}H)_{n-1}\ket{\phi(P_{2})}_{0,n-1},
    \vspace{10pt}
\end{equation}
where $H,X$ are the Hadamard and Pauli $X$ gates respectively. At a given number of intermediate qubits $n-2$, the teleportation produces exactly a total of four configurations of local transformations on $\ket{\phi(P_{2})}_{0,n-1}$ depending on the measurement results $\vec{s}$. Hence, using the identities $H^2=I$, $HX^{s_{i}}H=Z^{s_{i}}$ and ignoring global phase induced when swapping the order of $X$ and $Z$, we can manipulate \Cref{eq: local transformations_1} to obtain
\vspace{10pt}
\begin{equation}\label{eq: 4 configurations}
\begin{aligned}
    \ket{\psi}_{0,n-1}^{(\vec{s},n)}&=I_{0}\otimes(H^{n}Z^{s_{1}\oplus s_{3}\oplus\cdots}X^{s_{2}\oplus s_{4}\oplus\cdots})_{n-1}\ket{\phi(P_{2})}_{0,n-1}\\
    &\coloneqq\ket{\phi(P_{2})}^{(\vec{s},n)}_{0,n-1},
\end{aligned}
\vspace{10pt}
\end{equation}
where the symbol $\oplus$ denotes binary addition (XOR). No matter how long the path is, the teleported state $\ket{\psi}_{0,n-1}^{(\vec{s},n)}=\ket{\phi(P_{2})}_{0,n-1}^{(\vec{s},n)}$ always falls into one of the four configurations, and we can distinguish them by evaluating the discriminator inspired from \cite{discriminant_vec},
\vspace{5pt}
\begin{equation}\label{eq: categorisation}
\begin{aligned}
P(\vec{s})=\left(\bigoplus\limits_{i\in\text{odd}}{s_{i}},\bigoplus\limits_{j\in\text{even}}{s_{j}}\right).
\end{aligned}
\end{equation}
As a function of $\vec{s}$, the discriminator evaluates the powers of $X$ and $Z$ in \Cref{eq: 4 configurations} and the output for each is either 0 or 1. Therefore, there are precisely four possible outcomes of $P(\vec{s})$, each categorising one configuration of local transformation.

\section{Measures of teleportation success}\label{sec: Entanglement Measurements}
To determine the success of teleportation, we consider two properties in our 
demonstrations: negativity and fidelity.
\subsection{Negativity}
To quantify the amount of entanglement between two qubits of the teleported graph state as a Bell state up to local transformation in \Cref{eq: 4 configurations}, we employ an entanglement measure by calculating the negativity of the reduced density matrix of qubit 0 and qubit $n-1$ \cite{Negativity}. For a given density matrix $\rho$  of a quantum system, its negativity $\mathcal{N}(\rho)$ with respect to fixed bipartitions $a$ and $b$ is given by
\begin{equation}\label{eq: negativity}
    \mathcal{N}(\rho)=\frac{1}{2}(||\rho^{T_{a}}||-1)=\abs{\sum\limits_{\lambda_{i}<0}{\lambda_{i}}},
\end{equation}
where $\lambda_{i}$ are the negative eigenvalues of the partial transpose $\rho^{T_{a}}$ with respect to partition $a$ \cite{partial_transpose}. In our case, $a=0$ and $b=n-1$. The negativity has a value between 0 and 0.5, i.e. $0\le\mathcal{N}\le0.5$, and the higher the value implies more entanglement. N.B. for two qubits, non-zero negativity is a necessary and sufficient condition for entanglement \cite{bipartite_65}. 

To reconstruct the density matrix $\rho$, we perform full Quantum State Tomography (QST) on the resulting two-qubit state \cite{nielsen_chuang_2022}. The sampled probabilities for measurement in each two-qubit Pauli bases are calibrated with Quantum Readout Error Mitigation (QREM) \cite{QREM} to adjust for classical errors introduced during the qubit readout process. However, instead of calibrating all qubit readouts simultaneously, we correct the probability vector qubit-wisely \cite{Mooney_2021} such that the dimension of the calibration matrices only needs to be $2\times2$ for each iteration. In future work, an alternative method \cite{bo_yang_qrem} that is more efficient, could also be used. Unphysical negative probabilities can arise due to shot noise and device drift between the QREM calibration and the 
actual circuit execution. These are addressed using the Michelot algorithm to efficiently find the closest probability vector \cite{closest_pvec}. The constructed density matrix could contain unphysical negative eigenvalues caused by noise in the quantum computation. We use the efficient algorithm by Smolin et al. to find the nearest physical density matrix \cite{closest_physical_rhos}. 

Here we use the negativity to benchmark the entanglement preservation through teleportation by preparing the maximally entangled initial state $\ket{\phi(P_{2})}$. For a complete analysis that also considers potential bias in the initial state, we could further evaluate the process fidelity of teleportation. However, the additional circuits required to perform process tomography currently exceed our quantum resource allocation.

\subsection{Fidelity}
The fidelity of a measured state is its overlap with the expected ideal state, making it a natural measure for the success of teleportation. The fidelity \textit{F} of any prepared (noisy) state $\rho$ with respect to its ideal target state $\rho^{\text{ideal}}$ is defined as
\begin{equation}
    \textit{F}\left( \rho,{{\rho }^{\text{ideal}}} \right)=\text{tr}\left( \rho\rho^{\text{ideal}} \right).
\end{equation}
In the teleportation protocols, the state $\rho$, is the teleported 2-qubit graph state. It's density matrix is constructed using QST in the same manner as for measuring the negativity with QREM applied.

\section{Teleportation path optimisation}\label{sec: Teleportation path optimisation}
We propose three pathfinding protocols to maximise the amount of entanglement preserved after teleportation and the number of intermediate qubits that the state $\ket{\phi}$ can teleport across/displaced along the qubit layout of the quantum processor while maintaining entanglement, certified by non-zero negativity \cite{Negativity}. These protocols select the optimal path for teleportation from the quantum device based on the nearest-neighbour qubit-pair properties: entanglement negativity, negativity with quantum readout error mitigation (QREM), and two-qubit gate error. For a quantum device, we first construct an entanglement (or fidelity) graph $G$ where the vertices $V(G)$ represent all qubits and edges $E(G)$ represent the available couplings between them with weights assigned as the two-qubit negativity (or fidelity). The optimal $n$-qubit path $P^{(\text{opt})}_{n}\subseteq G$ used for teleportation is chosen to satisfy the following condition
\begin{equation}\label{eq: optimal path}
    \prod\limits_{e_{ij}\in E(P^{(\text{opt})}_{n})}{w(e_{ij})}=\max_{P'_{n}\subseteq G}\Biggl\{\prod\limits_{e'_{ij}\in E(P'_{n})}{w(e'_{ij})}\Biggr\},
\end{equation}
where $P'_{n}=(v_1,\cdots,v_n)$ is a path subgraph of $G$ with length $n-1$ also satisfying $d_{G}(v_1,v_n)=n-1$, which is the distance between nodes $v_1$, $v_n$ with respect to the original graph $G$. $w(e_{ij})$ is the weight of edge $e_{ij}$ assigned based on the pathfinding protocol. In the 
actual demonstrations, we select the best four paths (including $P^{(\text{opt})}_{n}$) based on the ordering of edge weight products using \Cref{alg: optimal paths}.

In the first protocol, we determine the optimal path $P^{(\text{neg graph opt})}_{n}$ at a given path length using the edge weight assignment $w(e_{ij})=2\mathcal{N}(\rho^{\text{Bell}}_{i,j})$, where $\mathcal{N}(\rho^{\text{Bell}}_{i,j})$ is the average negativity of the Bell states projected from the native graph state prepared on the whole device to the incident qubits $i,j$ via neighbouring qubit $Z$-basis projections \cite{fidel_paper}. The reason that the negativity is doubled is to expand the edge weights to the range of $[0,1]$ so that $w(e_{ij})=1$ indicates maximal entanglement. 

\begin{figure*}[t]
     \centering
     \includegraphics[width=0.85\linewidth]{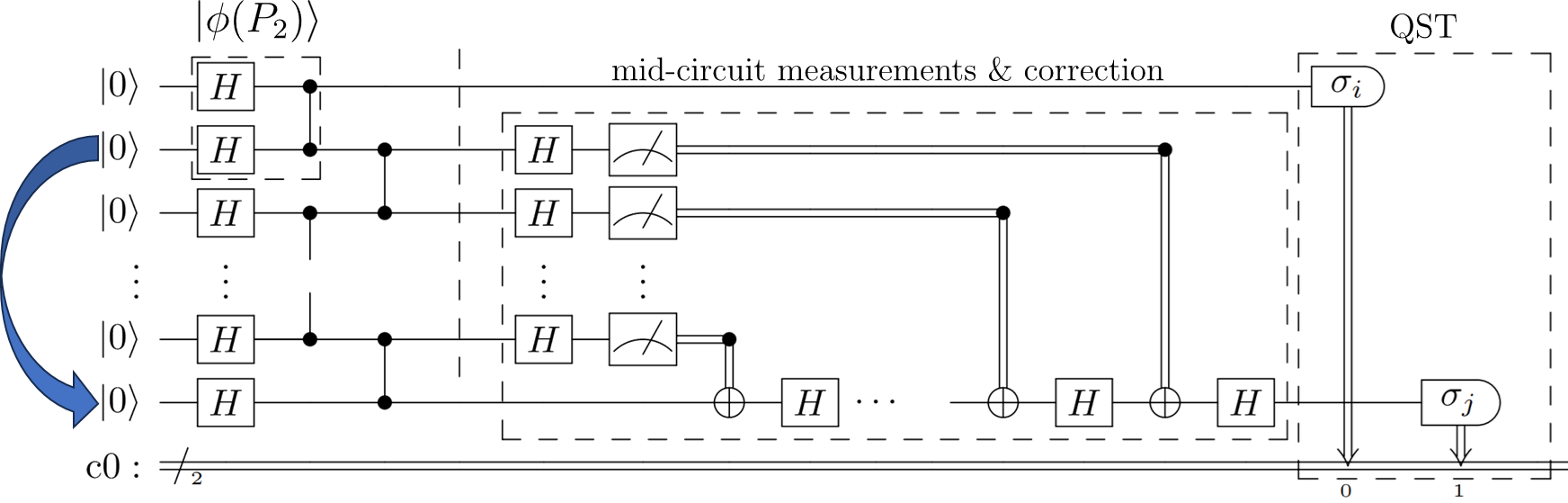}
     \caption{Illustration of the dynamic circuit approach to quantum teleportation. Initially, a graph state of $n$ qubits along the path $\ket{\phi(P_{n})}$ is generated. The intermediate qubits are measured in the Pauli-$X$ basis by applying a Hadamard gate before measuring in the computational Pauli-$Z$ basis. According to the measurement results, a sequence of $X$ and Hadamard gates are conditionally applied to the last qubit to undo the local transformation. At this stage, the two qubits of the two-qubit graph state $\ket{\phi(P_{2})}$ are now separated to the ends, as indicated by the blue arrow, making the overall quantum state in the ideal case equal to $\ket{\phi(P_{2})}\otimes\ket{s_{1}s_{2}\cdots s_{n-2}}$. Finally, Quantum State Tomography is performed to obtain the teleported two-qubit density matrix $\rho$ by measuring the first and last qubits in all combinations of Pauli bases.
     The double line at the bottom represents classical registers storing measurement results. 
     \label{fig:dynamic_circuit}}
\end{figure*}

In the second pathfinding protocol, we determine the optimal path $P^{(\text{neg QREM graph opt})}_{n}$ using the same assignment of edge weights as for selecting the optimal path $P^{(\text{neg graph opt})}_{n}$, except that $\mathcal{N}(\rho^{\text{Bell}}_{i,j})$ are measured using QREM.

In the third protocol, the optimal path $P^{(\text{weighted gate fid graph opt})}_{n}$ is chosen by assigning the edge weights to be the two-qubit gate fidelities determined from the IBM Quantum device gate-error map, using the expression
\begin{equation}
    w(e_{ij})=2\left(\frac{1}{2}-\epsilon_{ij}\right),
\end{equation}
where $\epsilon_{ij}\in\left[0,1/2\right]$ is the two-qubit gate error rate between qubits $i$ and $j$. Sometimes error rates calculated during IBM Quantum device calibrations are undefined and consequently displayed as a value of 1. Since error rates above 0.5 have no practical significance, they are excluded from the candidates of optimal paths. In our 
case, only 4 edges out of 144 are excluded, which has little effect on our choice of optimal path.

\section{Results}
Following the procedures on path optimisation, teleportation and entanglement measurement summarised in \Cref{fig:working flow chart}, we describe and implement two approaches to teleportation on the IBM Quantum \textit{ibm\_sherbrooke} device in \Cref{sec: dynamic circuits} and \Cref{sec: post-selected categorisation} respectively. We examine their performance using each of the path optimisation procedures, with and without QREM applied to the teleportation measurements. 
Optimal path scores calculated using the VF2 algorithm from the Python package \texttt{mapomatic} \cite{VF2++, mapomatic} are also compared with results shown in \Cref{appendix D}. Additionally, the performance of the teleportation protocols is compared with general state transportation using SWAP gates in \Cref{sec:comparisons}. To minimise the drift of noise parameters on the device, all circuits are completed within 10 hours with longest one no more than 24 hours.
The source code and datasets generated for this program are available under the GitHub repository \cite{Program_Codes}. The details of the quantum computer \textit{ibm\_sherbrooke} parameters calibrated at the times of execution are summarised as a table in the Supplementary Materials \cite{Sup_materials}.

\begin{figure*}[hbtp]
    \subfloat[\label{fig:dynamic_circuit_17gaps_unmitigated}Dynamic circuit two-qubit state negativity]{\hspace{-7pt}\:\includegraphics[width=1\columnwidth]{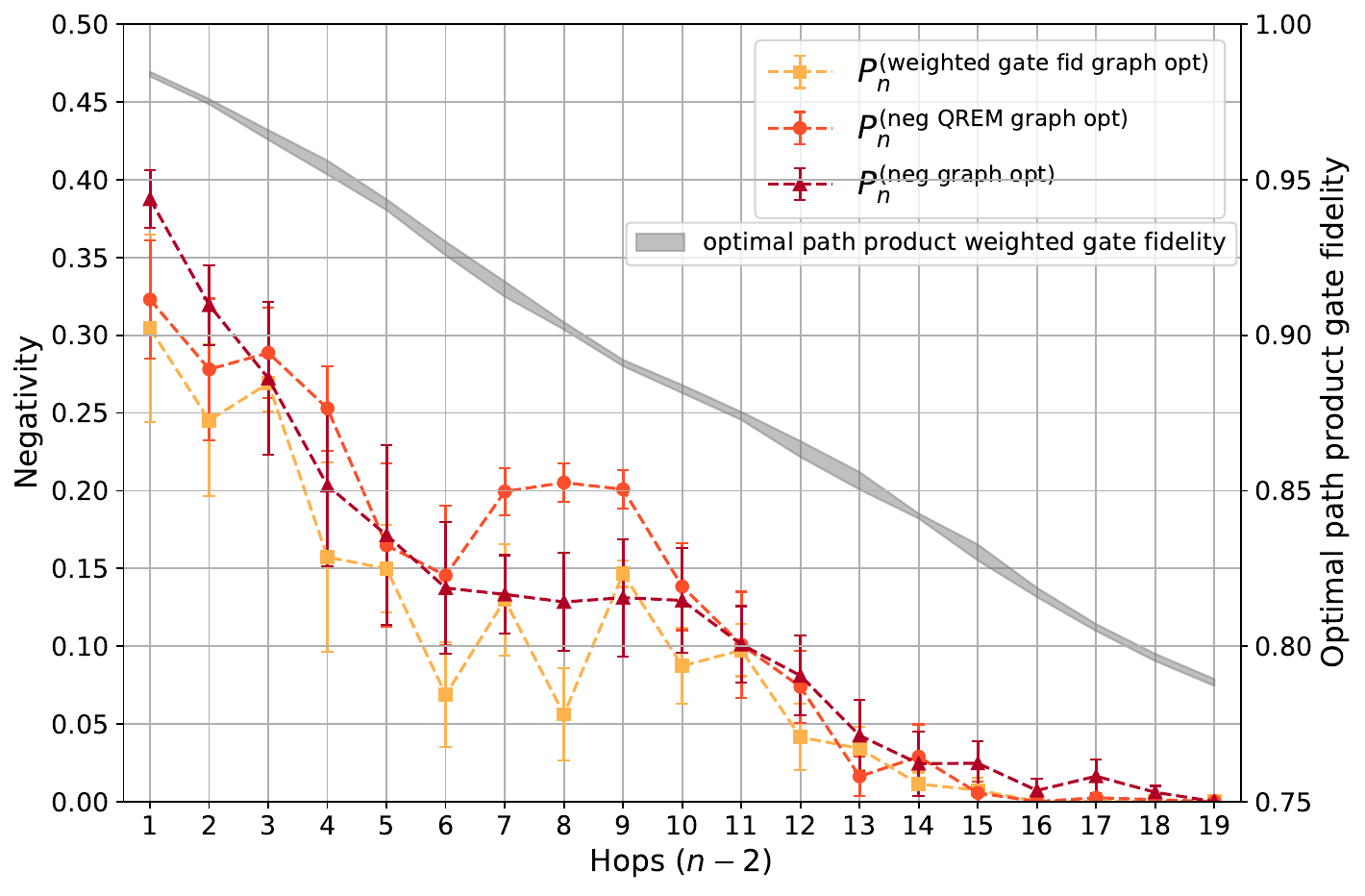}\:}
    \subfloat[\label{fig:dynamic_circuit_17gaps}Dynamic circuit two-qubit state negativity (QREM)]{\:\includegraphics[width=1\columnwidth]{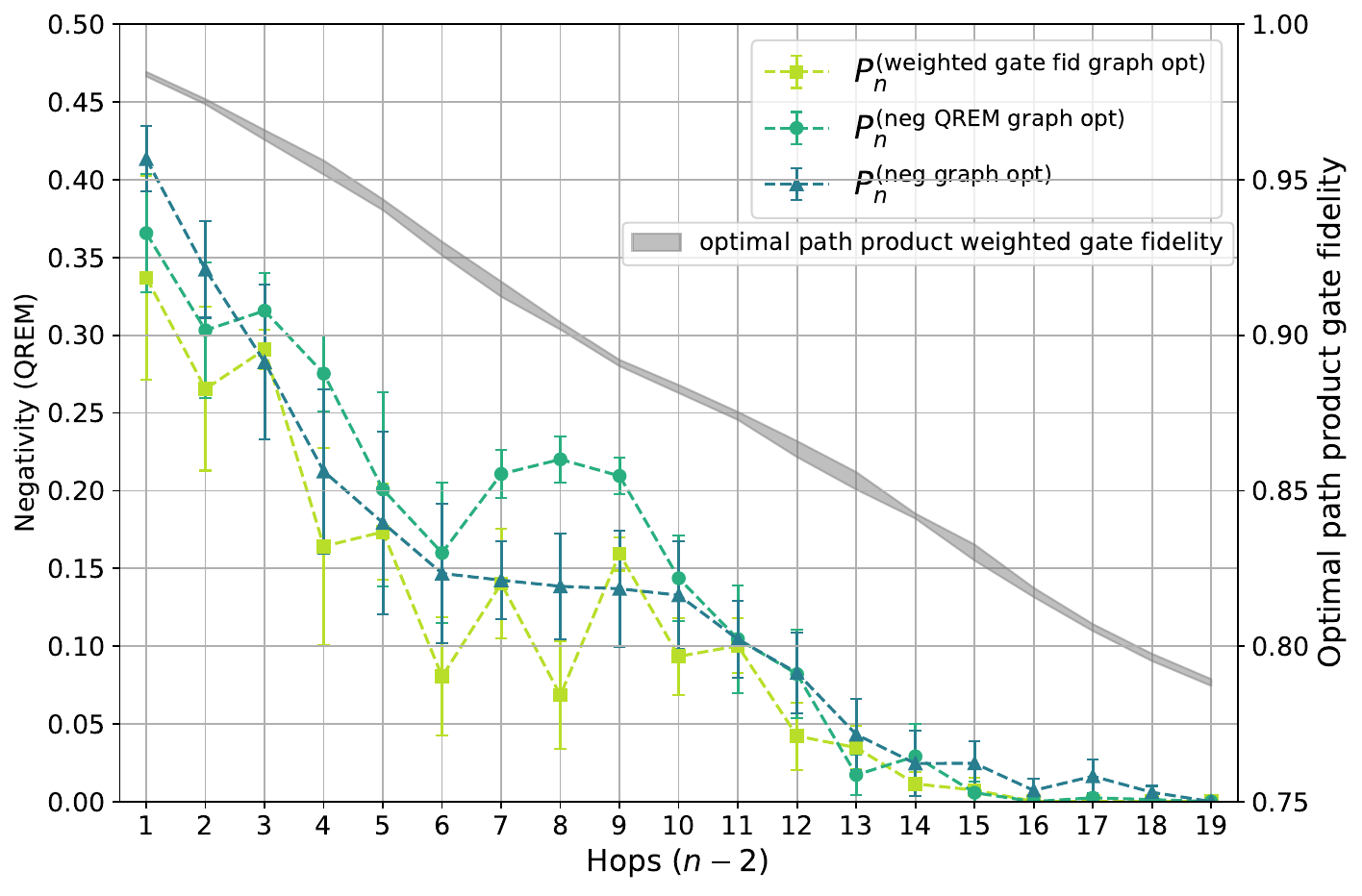}\:}\\
    \vspace{-10pt}
    \subfloat[\label{fig:dynamic_circuit_17_gaps_paths}]{\:\includegraphics[width=1.23\columnwidth]{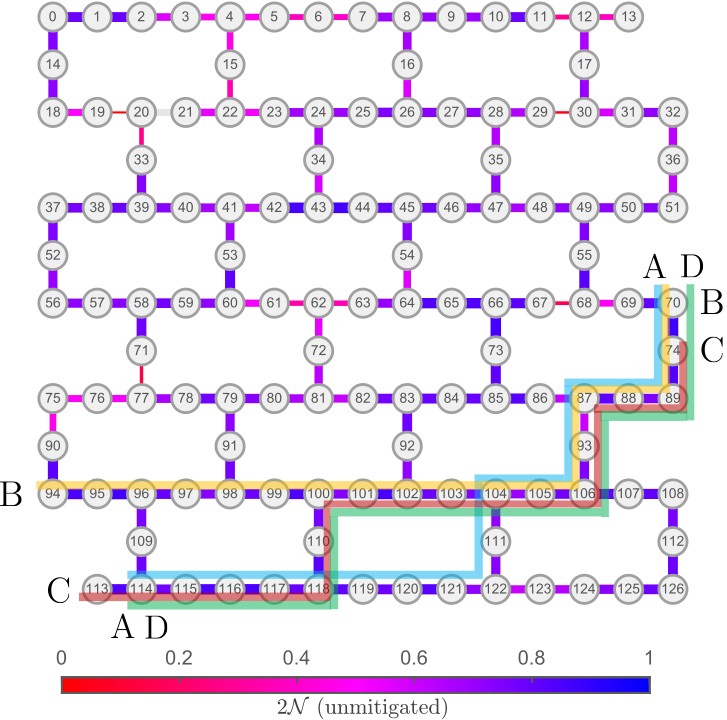}\:}
    \caption{\label{fig:dynamic circuit plots}
    Graphics showing results for the dynamic circuit approach to teleportation on optimal paths selected from the $ibm\_sherbrooke$ device. \textbf{(a)} The negativity without QREM measured on final two-qubit states against the number of qubits hops in teleportation. \textbf{(b)} The negativity with QREM measured on final two-qubit states against the number of qubit hops in teleportation. Due to controlled quantum operations after measuring intermediate qubits, QREM is naturally only applied to the two qubits. Both \textbf{(a)} and \textbf{(b)} also show the products of two-qubit weighted gate fidelities in $\{P^{(\text{weighted gate fid graph opt})}_{n}\}$ on the second vertical axis as a point of reference. The different optimal path protocols used are represented by distinct colours on the plot, where each data point has its negativity averaged over the best four paths at the given number of hops and each path is sampled in four trials. Error bars use two times the standard error of the sampled negativities. \textbf{(c)} The four paths found with the highest net product of nearest-neighbour negativity $\{P^{(\text{neg graph opt})}_{19}\}$ as described in \Cref{sec: Teleportation path optimisation} that have a path length of 18 (17 intermediate qubits). The paths are drawn on top of the coupling map, where each node represents a qubit and coloured edges represent the available nearest-neighbour couplings with weights $w(e_{ij})=2\mathcal{N}_{ij}$, where $\mathcal{N}_{ij}$ is the negativity measured between qubits $i$ and $j$. Each path is displayed with a distinct colour, and its ends are marked by the same letter: A, B, C, D.}
\end{figure*}

\subsection{Dynamic circuit approach to teleportation}\label{sec: dynamic circuits}
We perform the measurement-based teleportation protocol using the IBM Quantum dynamical circuits capability. Similarly to the principle behind the entanglement swapping \cite{entanglement_swapping}, mid-circuit measurements are performed on entangled states followed by byproduct correction operators. Since measurement gates are inserted into the middle of a quantum circuit, this results in a collapse of the quantum state that is not recoverable by applying a set of Hermitian conjugate gates \cite{one_way_qc}. However, this also enables dynamic modification of the circuit during execution by directly measuring some of its qubits and conditionally applying gates to the rest of the circuit based on the outcomes. We note that one-way quantum computation can be performed on IBM Quantum devices through the use of dynamic circuits \cite{detect_entanglement_via_teleportation, long_range_gates}. In this subsection, we will discuss the use of byproduct operators for correction after mid-circuit measurement within the teleportation procedure. 

Using the procedure outlined in \Cref{sec: Teleportation}, we first design the circuits with dynamical modifications to measure the negativities of the two-qubit graph states after teleportation. As in \Cref{fig:dynamic_circuit}, we prepare the graph state on a path of $n$ vertices and measure intermediate qubits in the Pauli-$X$ basis by applying a Hadamard gate before measuring in the computational Pauli-$Z$ basis. To recover the form of the initial two-qubit graph state, $\ket{\phi(P_{2})}$, dynamical circuits are used to apply byproduct operators on the teleported two-qubit state, where a set of conditional Pauli-$X$ and Hadamard gates are sequentially acted on the last qubit. The purpose of implementing byproduct operators after mid-circuit measurements is to undo any variations of the state caused by projections from measured intermediate qubits. Ideally, the sequence of gates can be simplified to a maximum of two sequential Clifford gates using the discriminant vector shown in \Cref{eq: categorisation}, ensuring the number of quantum gates involved in the correction remains constant with respect to the number of qubits. However, the IBM Quantum devices do not currently support a classical XOR operation acting on the classical register required for the simplification. Although a classical switch could be used as a workaround, however, to the best of our knowledge, it is only feasible on short paths since it would require a case for every outcome bit-string in the control flow, which in principle scales exponentially. Thus, all byproduct operators are applied sequentially. Note that during the application of byproduct operators, instead of applying quantum CNOT gates to the last qubit controlled on post-measured qubits, IBM Quantum devices actively apply quantum $X$ gates based on the mid-circuit measurement results. The classical layer of processing involved in dynamic circuits includes latency overhead that limits the readout and write speeds.

In \Cref{fig:dynamic circuit plots}, we present the results for the dynamic circuit approach to teleportation on optimal paths selected from the $ibm\_sherbrooke$ device. \Cref{fig:dynamic_circuit_17gaps_unmitigated} and \ref{fig:dynamic_circuit_17gaps} show the negativity of the final two-qubit state against the number of qubit hops in teleportation for the cases of no QREM and QREM respectively. Negativity is derived from the density matrix of the two-qubit graph state teleported along optimal paths found using the protocols outlined in \Cref{sec: Teleportation path optimisation} with varying numbers of intermediate qubits. As we increment the number of teleportation hops, the average measured negativity of the final teleported state decreases. Although statistically, it is not significant that the negativity is nonzero at 16 hops since 0 is included in the errorbar range, the maximal number of teleportations yielding reliable non-zero negativity is 17, with the best four paths $\{P^{\text{(neg graph opt)}}_{19}\}$ shown in \Cref{fig:dynamic_circuit_17_gaps_paths}. This apparent discrepancy is likely due to statistical variation and fluctuations in device parameters over time. However, due to the dynamic corrections to the teleported state, the first qubit in the teleported state remains idle during classical processing. This leads to a rapid overall decrease in negativity caused by decoherence and free rotations \cite{fidel_paper}, significantly impacting the level of entanglement during teleportation. To help alleviate the problem of idling qubits, one could implement dynamical decoupling techniques to minimise decoherence \cite{dynamical_decoupling}.  
Another observation is that regardless of whether QREM is applied, the data show an improvement in teleported entanglement on average, although modest, when states are teleported along the paths $\{P^{\text{(neg graph opt)}}_{n}\}$ and $\{P^{\text{(neg QREM graph opt)}}_{n}\}$ compared to $\{P^{\text{(weighted gate fid graph opt)}}_{n}\}$. These results indicate an advantage in using the negativity map over the gate error map for finding optimal paths, highlighting its potential in general quantum circuit compilation. 
Furthermore, when \texttt{mapomatic} is used to calculate cost function scores for $P_n^{\text{(weighted gate fid graph opt)}}$ in \Cref{fig:Scores mapomatic} of \Cref{appendix D}, they are found to be smaller than the costs for $P_n^{\text{(neg graph opt)}}$ and $P_n^{\text{(neg QREM graph opt)}}$, however, our 
result in \Cref{fig:dynamic circuit plots} shows the opposite. The discrepancy between what one might expect in \Cref{fig:Scores mapomatic} and the actual sampled results of Figures \ref{fig:dynamic circuit plots} and \ref{fig: fidelity plots detailed} suggests that the cost functions based on negativity provide a potential improvement over the \texttt{mapomatic} algorithm, although future investigation is required. 


\raggedbottom

\begin{figure*}[hbtp]
     \centering
     \includegraphics[width=0.72\linewidth]{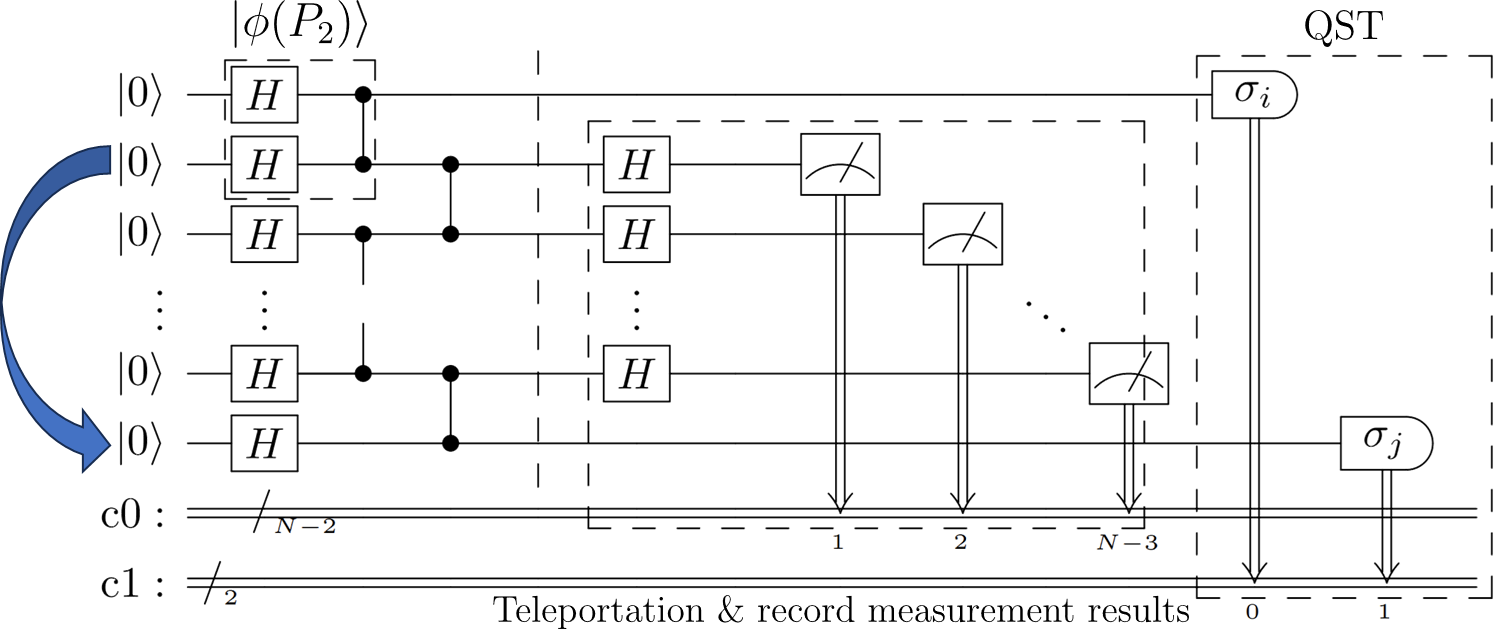}
     \caption[]{Illustration of the teleportation circuit based on post-selection. The key distinction from the protocol outlined in \Cref{fig:dynamic_circuit} lies in the absence of conditional modifications during circuit execution. Instead, we categorise the projected state through classical post-selection. Thus, the exact local transformations on the teleported two-qubit graph state are identified according to the measurement outcomes stored in the classical register $c0$ for each shot. If ignoring the measurement results of intermediate qubits, the system exists in an overall mixed state with pure state components of $\ket{\phi(P_{2})}^{(\vec{s},n)}\otimes\ket{s_{1}s_{2}\cdots s_{n-2}}$ before QST.
     \label{fig:post processed circuit}}
\end{figure*}

Additionally, there appears to be a considerable improvement for paths with 3 to 9 qubit hops when chosen from $\{P^{\text{(neg QREM graph opt)}}_{n}\}$, especially from 6 to 9 hops which are more statistically significant. This behaviour is likely due to the two methods of measuring negativities weighting sources of noise differently. In particular, decoherence and 2-qubit free rotations \cite{fidel_paper} are likely to be more emphasised when choosing optimal paths if QREM is applied due to the reduced impact of readout errors. Since dynamical circuits involve long idle times for classical processing, this emphasis on non-readout related errors achieved with QREM could be beneficial. Longer paths naturally have higher levels of readout error. Thus, the performance of paths chosen without QREM $\{P^{\text{(neg graph opt)}}_{n}\}$, which put more weight on readout errors, would be expected to improve over $\{P^{\text{(neg QREM graph opt)}}_{n}\}$. This could explain their slight advantage for paths with 15 or more hops. The corresponding fidelity versus number of hops plots are included in \Cref{appendix C}. To further account for the impact of decoherence due to long idle times, the characteristic amplitude damping ($T_1$) and the dephasing time ($T_2$) could be included in the cost function when finding optimal paths. To circumvent the long idle times involved in the classical processing of mid-circuit measurements, we investigate post-selection based teleportation in the next subsection. 

\begin{figure*}[hbtp]
    \subfloat[\label{fig:post processed circuit 19gaps unmitigated}Post-selected two-qubit state negativity]{\hspace{-7pt}\:\includegraphics[width=1\columnwidth]{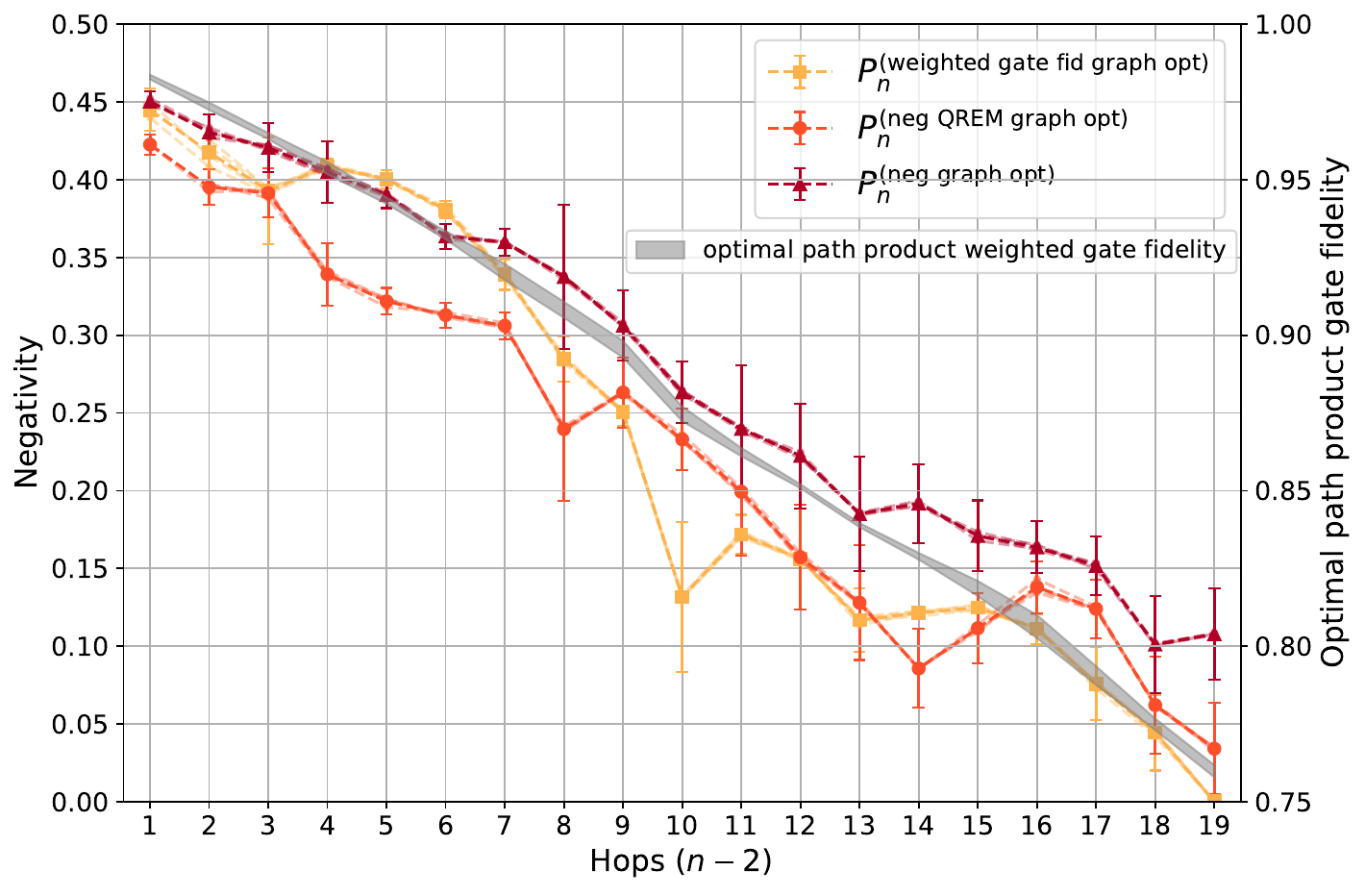}\:}
    \subfloat[\label{fig:post processed circuit 19gaps}Post-selected two-qubit state negativity (QREM)]{\:\includegraphics[width=1\columnwidth]{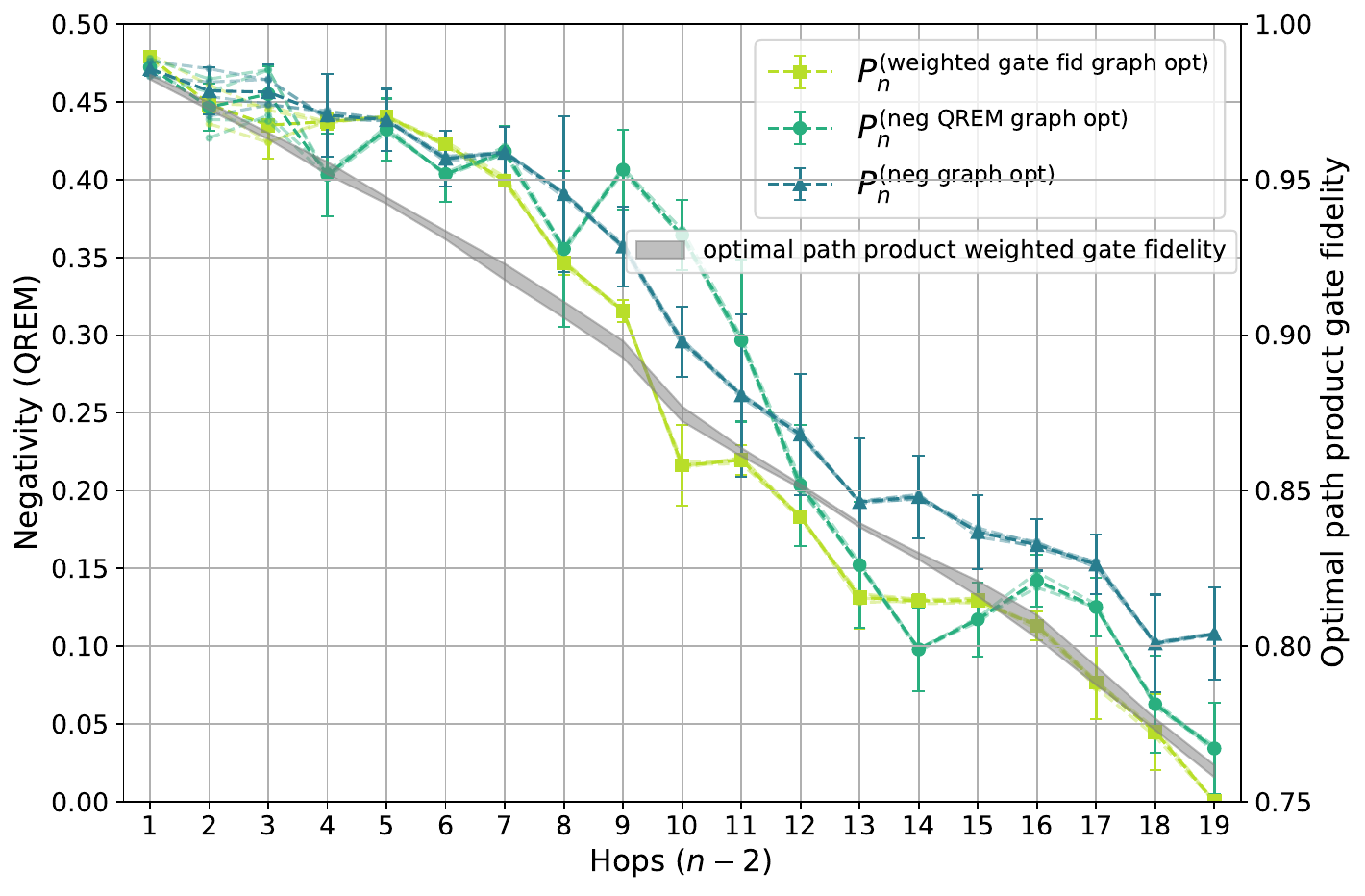}\:}\\
    \vspace{-10pt}
    \subfloat[\label{fig:post_processed_19_gaps_paths}]{\:\includegraphics[width=1.23\columnwidth]{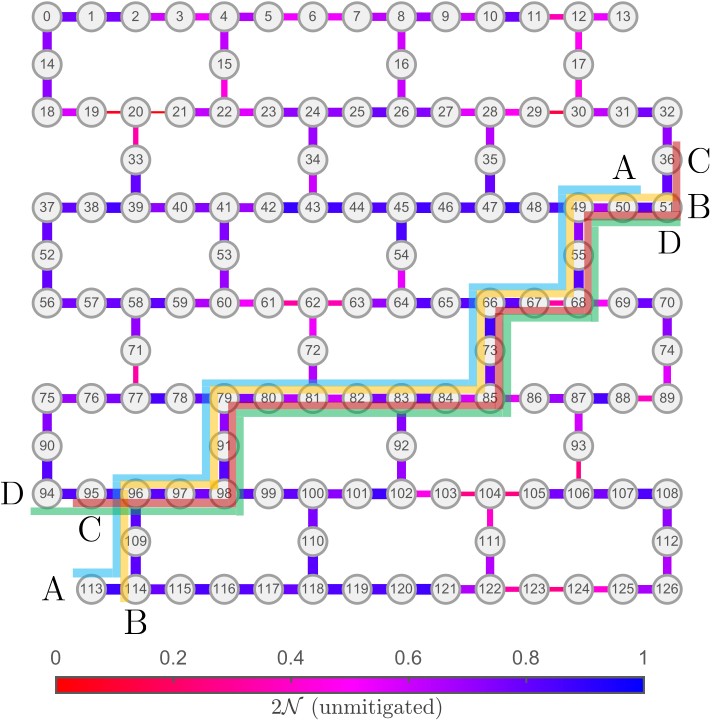}\:}
    \caption{\label{fig:post processed plots} Graphics showing results for the post-selected approach to teleportation on optimal paths selected from the \textit{ibm\_sherbrooke} device. \textbf{(a)} The negativity without QREM measured on final two-qubit states against the number of qubits hops in teleportation. \textbf{(b)} The negativity with qubit-wisely applied QREM measured on final two-qubit states against the number of qubits hops in teleportation. Both \textbf{(a)} and \textbf{(b)} show the products of two-qubit weighted gate fidelity on $\{P^{(\text{weighted gate fid graph opt})}_n\}$ on the second vertical axis as a point of reference. The different optimal path protocols used are represented by distinct colours on the plot, where each data point has its negativity averaged over the best four paths at the given number of hops and each path is also sampled in four trials. Different from \Cref{fig:dynamic circuit plots}, there are also light-coloured dashed lines to distinguish the results among four configurations. N.B. that only two configurations are available when there is just one hop, and the remaining two are omitted in the plot. Error bars represent two times the standard error of the sampled negativities. \textbf{(c)} The four paths found with the highest net product of nearest-neighbour negativity $\{P^{(\text{neg graph opt})}_{21}\}$ described in \Cref{sec: Teleportation path optimisation} with a path length of 20 (19 intermediate qubits). The paths are drawn on top of the negativity coupling map of the quantum device $\textit{ibm\_sherbrooke}$, where each node represents a qubit and coloured edges represent the available nearest-neighbour couplings with weights $w(e_{ij})=2\mathcal{N}_{ij}$, where $\mathcal{N}_{ij}$ is the negativity measured between qubits $i$ and $j$. Each path is displayed with a distinct colour, and its ends are marked by the same letter: A, B, C, D.}
\end{figure*}
\begin{figure*}[!btp]
     \centering
     \includegraphics[width=0.5\linewidth]{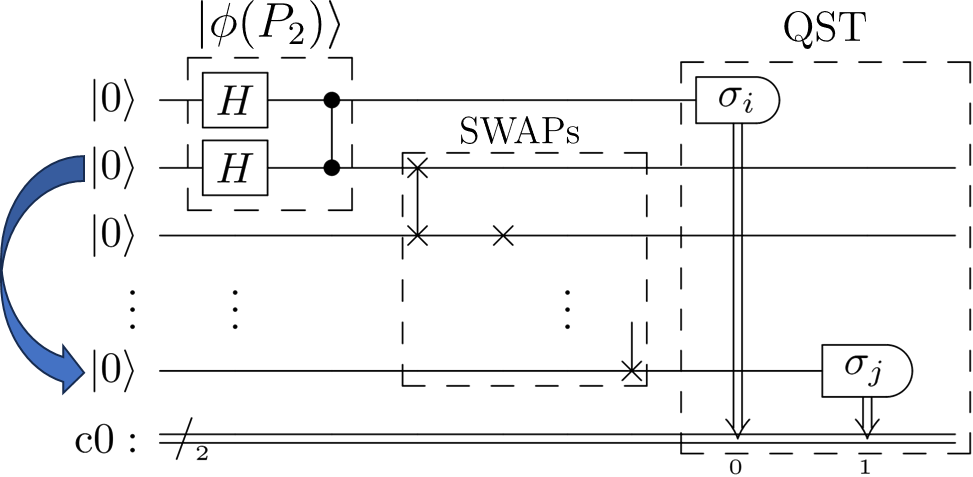}
     \caption[]{Circuit diagram of the simple and naive design to transport the information of the second qubit to the last by continually applying SWAP gates. Ideally, the quantum state before QST should be in the form of $\ket{\phi(P_{2})}_{0,n-1}\otimes\ket{00\cdots0}$.
     \label{fig:SWAP circuit}}
\end{figure*}

\subsection{Post-Selected approach to teleportation}\label{sec: post-selected categorisation}
Similar to the mid-circuit measurements based teleportation, we also employ $X$-basis measurements to project the two-qubit graph state. However, instead of inserting quantum gates dynamically after measurements, we categorise the teleported state into one of the four possible configurations using the discriminator shown in \Cref{eq: categorisation} on the measurement outcomes of intermediate qubits. This approach aims to tackle the problem of errors arising due to idling qubits during mid-circuit measurements by reducing the circuit depth and shifting measurements to the end of the circuits. \Cref{fig:post processed circuit} outlines the modified algorithm. By recording the measurement outcomes of the intermediate qubits, we identify the teleported state configuration after the circuit is completed. Since the resulting two-qubit state can be categorised directly by \Cref{eq: categorisation}, there is no need to apply byproduct operators, and the entanglement for each category can be calculated independently. Therefore, the depth of the quantum circuit is reduced and the mid-circuit measurements are no longer needed. This largely avoids errors caused by decoherence and free rotations, establishing a cleaner environment for benchmarking the loss of information caused by gate noise during teleportation. By its nature, the post-selection approach acts as an upper bound to the dynamic circuit approach, hence its performance is expected to be higher. It also enables us to apply QREM on the full path of qubits. However, the technique of qubit-wise QREM has its accuracy limited by the zero-out threshold, especially when the outcome probabilities dissipate into the whole outcome space during QREM which scales exponentially. Hence, we set a small threshold (0.001/number of shots) to minimise such an effect. 

Unlike the correction after mid-circuit measurements, post-selection would project the teleported state into one of the four variants randomly at a fixed number of intermediate qubits previously stated in \Cref{sec: Teleportation}. Reconstructing the density matrix $\rho=\ket{\phi(P_{2})}\bra{\phi(P_{2})}$ directly would obtain a mixed state,
\begin{equation}
    \rho^{(\text{mixed})}=\frac{1}{2^{n-2}}\sum\limits_{\vec{s}}{\ket{\phi(P_{2})}^{(\vec{s},n)}\bra{\phi(P_{2})}^{(\vec{s},n)}}.
\end{equation}
Instead, the measurement outcomes are categorised into the four possible Bell state/two-qubit graph state configurations before calculating their corresponding density matrices $\rho^{(\vec{s},n)}$. Since the categorisation can be easily determined from the discriminant vector as shown in \Cref{eq: categorisation}, the process of undoing local transformations $S=(X^{s'_{n-2}}H)\cdots(X^{s'_{2}}H)(X^{s'_{1}}H)$ by computing $\rho=S^\dagger \rho' S$ is not necessary. 

\Cref{fig:post processed circuit 19gaps unmitigated} and \Cref{fig:post processed circuit 19gaps} show the average negativities measured from each of the four configurations on $ibm\_sherbrooke$ as a function of the number of hops teleported in light-coloured lines. The mean negativities averaged over four configurations are shown in dark-coloured lines. Similar to results from \Cref{fig:dynamic_circuit_17gaps_unmitigated}, \ref{fig:dynamic_circuit_17gaps}, the two-qubit graph state is teleported along the same optimal paths using the same protocols. We also observe the negativity decreases as the number of qubits hopped during teleportation increases, indicating non-perfect teleportations due to noise, as well as the improvement in the measured negativity after applying qubit-wise QREM. However, on average the negativity preserves better compared to the approach using mid-circuit measurements. Even after hopping 19 qubits, the entanglement of the teleported state recognised by post-selection still maintains significant non-zero negativity, whereas the negativities measured from the mid-circuit measurement approach all drop to zero. Specifically, the mitigated negativity of the states teleported along the best four paths $\{P^{(\text{neg graph opt})}_{21}\}$ (as shown in \Cref{fig:post_processed_19_gaps_paths}) has an average reading of 0.108. This strongly demonstrates the advantage of post-selection in preserving entanglement during teleportation by reducing circuit depth and shifting measurements to the end. The corresponding fidelity versus number of hops plots are included in \Cref{appendix C} that shares very similar pattern. We anticipate that the highest number of teleportation hops is not limited to 19, suggesting a direction in future research of adding more hops to teleportation until the extreme is reached. However, if one must insist on obtaining a specific projected state when running multiple teleportation circuits in parallel, the number of shots that the post-selection method requires scales exponentially with respect to the number of teleportation paths since the expected number of shots that projects to the correct state is only a quarter of the total. The number of shots does not scale with the number of qubits in the path, since the number of possible variants is always four (two if only one hop).

Compared to the results from mid-circuit measurement 
demonstrations, it is more evident that, on average the entanglement of the states teleported along $\{P^{(\text{neg graph opt}))}_{n}\}$ paths preserve more entanglement than $\{P^{(\text{weighted gate fid graph opt}))}_{n}\}$ both in \Cref{fig:post processed circuit 19gaps unmitigated} and \Cref{fig:post processed circuit 19gaps}. Additionally, more qubits are teleported across while the negativity remains non-zero (19 in $\{P^{(\text{neg graph opt}))}_{n}\}$ versus 18 in $\{P^{(\text{weighted gate fid graph opt}))}_{n}\}$). This further highlights the importance of entanglement benchmarking using nearest-neighbour negativity \cite{fidel_paper}, as a better indicator of optimal paths. However, as the number of intermediate qubits increases, despite the considerable amount of improvement in the measured negativity compared to \Cref{fig:post processed circuit 19gaps unmitigated}, we still find that the results from $\{P^{(\text{neg QREM graph opt}))}_{n}\}$ paths underperform the results from $\{P^{(\text{neg graph opt}))}_{n}\}$ paths for larger number of hops in \Cref{fig:post processed circuit 19gaps}, which also happens consistently in \Cref{fig:dynamic_circuit_17gaps}. This is likely due to the negativities calculated without QREM putting more weight on readout errors compared to with QREM, since longer chains become more dominated by readout errors, making negativity without QREM a better indicator for path optimisation. We also suspect that $\{P^{(\text{neg QREM graph opt})}_{n}\}$ could actually put more weight on qubits that benefit more from QREM, leading to QREM being more likely to be effective on the teleportation protocol as well. This could be investigated in future work.

\subsection{Comparisons with SWAP gates approach}\label{sec:comparisons}
\begin{figure}[t!]
     \centering
     \subfloat[\label{fig:teleportation comparisons}Negativity comparison between protocols\vspace{-5pt}]{
         \centering
         \hspace{-8pt}
         \includegraphics[width=0.925\linewidth]{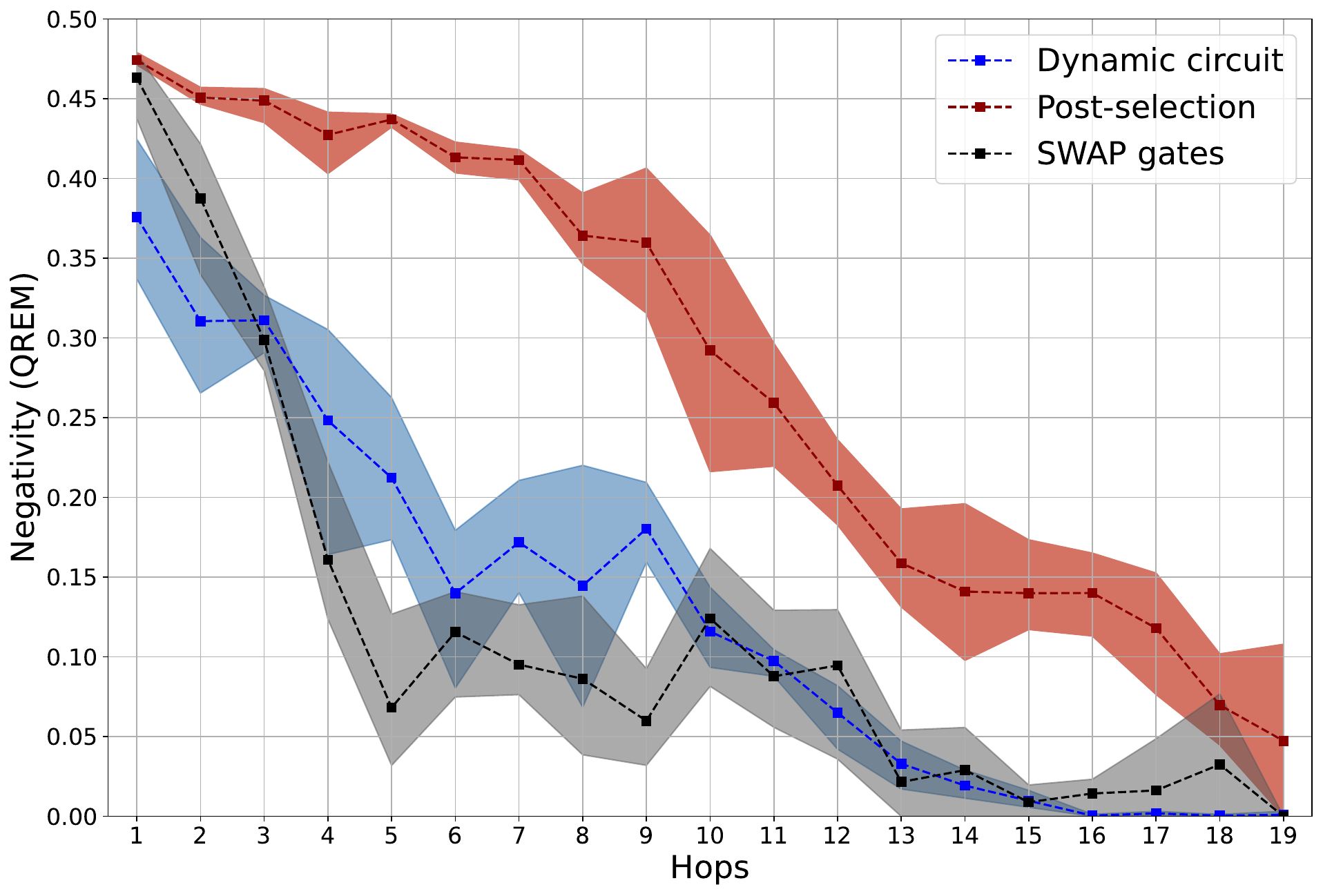}
     \vspace{-50pt}}
     \\
     \subfloat[\label{fig:teleportation comparisons fidelity}Fidelity comparison between protocols\vspace{-5pt}]{
         \centering
         \hspace{-8pt}
         \includegraphics[width=0.925\linewidth]{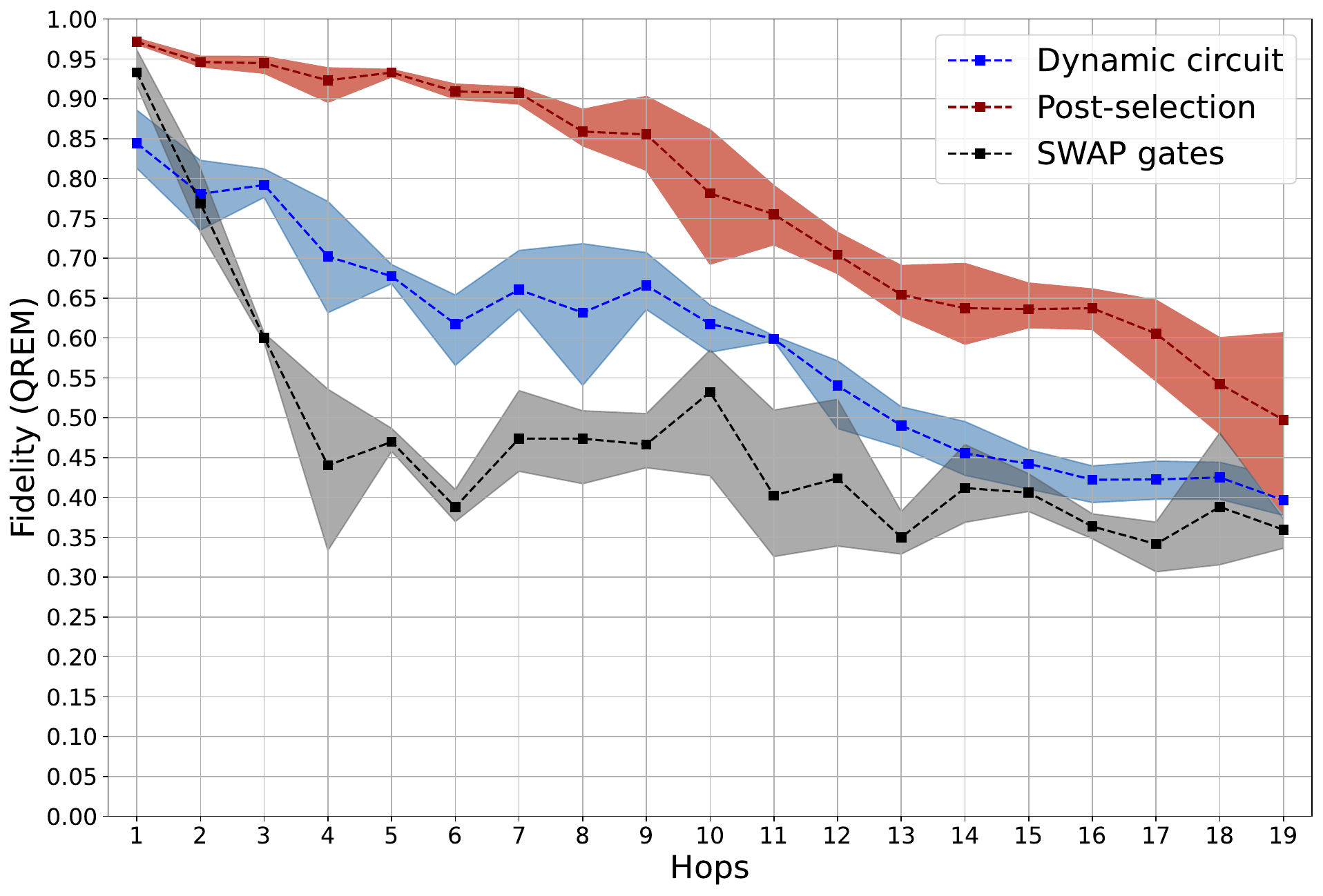}
     }
     \caption[]{\label{fig:teleportation negativity and fidelity}Plots that compare the performance between the two teleportation protocols and the SWAP gate approach.\\
     \textbf{(a)} QREM mitigated negativity measured using dynamic circuit, post-selected categorisation, and SWAP gates are plotted together. The shaded region represents the range of negativity measured from all path-optimising protocols; heavy dashed lines are highlighted for their means. \textbf{(b)} Same comparison on correction after mid-circuit measurements, post-selected local transformations, and SWAP gates, but using fidelity.\label{fig:comparisons}}
\end{figure}

In previous subsections, we have shown that the entanglement decays as the number of hops in teleportation grows. We also comment on the performance of post-selected categorisation that surpasses the performance of the mid-circuit measurement approach. As a point of comparison for the teleportation approaches, we investigate the performance of a naive implementation of the general SWAP gate approach for qubit transportation, which scales linearly in 2-qubit gates with respect to the number of qubits. The SWAP gate approach is illustrated in \Cref{fig:SWAP circuit}, where
\begin{equation}
    \text{SWAP}_{01}\ket{\alpha}_{0}\otimes\ket{\beta}_{1}=\ket{\beta}_{0}\otimes\ket{\alpha}_{1}
\end{equation}
simply swaps the quantum states between two qubits on the computational basis. We also compare the entanglement measured across all discussed approaches in \Cref{fig:comparisons}, emphasising the improvement achieved in post-selected categorisation.

In \Cref{fig:teleportation comparisons}, we plot the mitigated negativity averaged over all three protocols of optimal paths against the number of hops for both teleportation schemes and the results obtained using SWAP gates. Overall, the entanglement of the state teleported with post-selected categorisation consistently preserves better, even after considering all of the negativity measured regardless of the choice of paths as shown in the shaded region. This highlights the advantage of the post-selection method in reducing circuit depth and decoherence of idling qubits during dynamic operations. We also notice that on average, the post-selected categorisation method achieves the longest teleportation distance of 19 hops before the final negativity drops to zero, compared to the other methods that are capped at 17 hops. The faster reduction of entanglement in the graph states using SWAP gates is likely due to the large circuit depth that grows linearly with respect to the number of hops and contributes three CNOTs for every SWAP gate. In contrast, while preserving a relatively high amount of entanglement, the classical post-selected categorisation after teleportation does not rely on additional quantum gates to transmit messages like SWAP gates or byproduct operators.

To reinforce the results in \Cref{fig:teleportation comparisons} that compare the quality of entanglement using negativity, we also characterise the same teleported states using fidelity as described in \Cref{sec: Entanglement Measurements}, which tells us the overlap between a noisy and ideal state. In \Cref{fig:teleportation comparisons fidelity}, we observe similar behaviour of the teleported state as in \Cref{fig:teleportation comparisons}, including an overall trend of decaying fidelity as the number of teleportation hops increase, as well as significantly higher reading of the fidelity from post-selection method. 

Being a direct measure of quantum entanglement, the 2-qubit negativity is highly tied to the quantum coherence of the state, while the fidelity of the state with respect to an ideal 2-qubit maximally entangled state considers both its quantum and classical properties in relation to this reference state. It is possible for the state to simultaneously have a high amount of entanglement and a low amount of fidelity (e.g. when the state is rotated to be orthogonal to the ideal state). However, the fidelity is limited by the amount of entanglement because the negativity should approach its maximum value of 0.5 as the fidelity approaches 1 since the ideal state's negativity is 0.5. While the negativity in \Cref{fig:teleportation negativity and fidelity} decays towards 0, the fidelity does not decay asymptotically towards 0.25, which corresponds to the maximally mixed state. Instead, the measured fidelity is higher. This potentially indicates the existence of other noise channels during circuit execution that results from other noise channels such as crosstalks with environment. The detailed plots for each of the teleportation schemes and path-optimising protocols are shown in \Cref{appendix C}. Overall, the results indicate that we have preserved the entanglement of two-qubit states after teleporting them across at least 19 qubits, which benchmarks a high level of performance on IBM Quantum computers in quantum state teleportation.

\begin{figure}[t]
    \centering
    \captionsetup{width=1\linewidth}
    \includegraphics[scale=0.39]{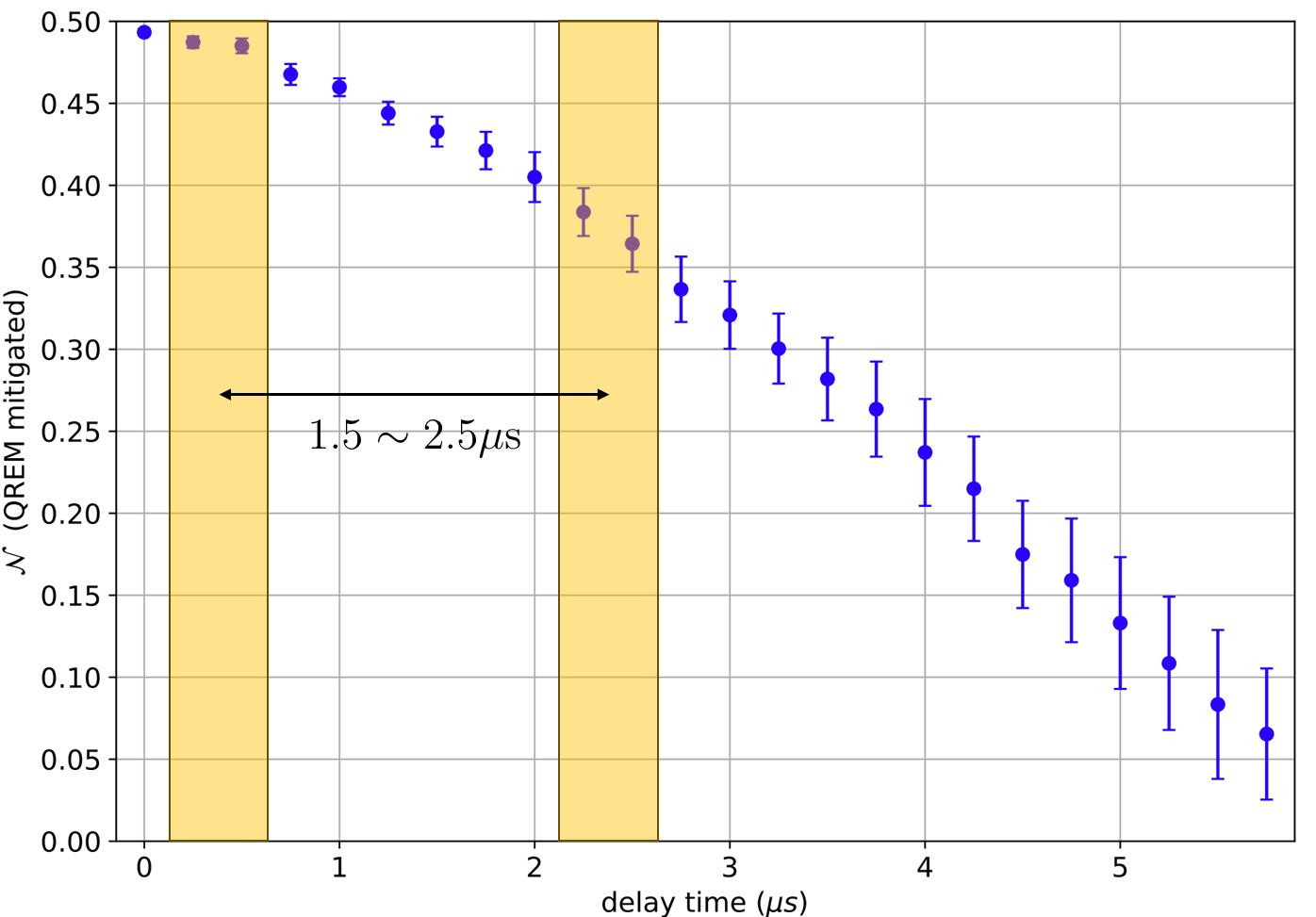}
    \caption{Plot of the average two-qubit graph state QREM mitigated negativity versus delay time on the IBM Quantum device $ibm\_sherbrooke$. The average time for the negativity to decay from near 0.474 (teleportation by once using post-selected categorisation with QREM, marked by the first band) to near 0.376 (teleportation by one hop using mid-circuit measurement with QREM, marked by the second band) is roughly 1.5\textmu s to 2.5\textmu s.}\vspace{-5pt}
    \label{fig:2q GS decay}
\end{figure}

In addition, we observe on average a considerable gap of 0.098 in the negativity between correction after mid-circuit measurement and the post-selected categorisation method even after just one hop. Yet the circuit's depth after applying the corrections is only increased by at most two. We attribute this to the long gate execution time of the dynamic CNOT gate, such that there is a significant delay in the circuit within just one hop. To quantify the delay, we measure the average negativity of the two-qubit graph state simply idling on the same machine in \Cref{fig:2q GS decay}. The time it takes for the negativity to reduce by the amount indicated above (from 0.474 to 0.376) is roughly 1.5-2.5\textmu s. A recent study performed on the same type of quantum processor \cite{feed_forward_time} showed that the time to complete mid-circuit correction for a single qubit is roughly 1954 ns (1244 ns for readout, 653 ns for feed-forward on $X$ gate, and 57 ns for $\sqrt{X}$ gate after compiled from the Hadamard gate) when measurement, feed-forward and all subsequent byproduct operator gate times are taken into account. This agrees with our estimation and the effective gate time of the dynamical equivalent to a CNOT gate, it is indeed much longer than the typical two-qubit gate time on $\textit{ibm\_sherbrooke}$, which has the mean of 533 ns. 

\section{Discussion}
In this paper, we have demonstrated the teleportation of two-qubit graph states along the path of entangled qubits on physical quantum computers. For comparison purposes, two protocols were used: strict teleportation using a dynamic circuit, and a version based on post-selection. These were compared directly with direct SWAP-based transportation. The results showed the method of post-selected categorisation
outperformed the others in terms of minimising information lost during teleportation hops. Under this approach,
we sustained the original state entanglement after teleporting 19 hops. We also compared the performance of different
optimal pathfinding protocols. In the scenario where the method of post-selected categorisation is carried out, we found that the path that maximises the product of negativities between nearest neighbours sustained entanglement the most, which reflects the utility of entanglement benchmarking using the nearest-neighbour negativity as opposed to the gate error map. In future work, it would be worth calculating the process fidelity of teleportation to account for potential bias in the choice of initial state.

This work provides 
verified tools to realise more advanced circuit compilation applications in future research. These include qubit pathfinding and the implementation of high-fidelity long-range gates that do not rely on SWAP gates, which are also important elements in creating large-scale entangled states and benchmarking the capabilities of quantum devices. Furthermore, the results potentially offer a scalable method of measuring entanglement on large, one-dimensional graph states.

\bigskip
\textbf{Acknowledgments}
\par
This work was supported by the University of Melbourne through the establishment of an IBM Quantum Network Hub at the University. The authors gratefully acknowledge the support of the University of Melbourne’s Zero Emission Energy Laboratory (ZEE Lab) and the Victorian Higher Education State Investment Fund (VHESIF). JFK was supported by an Australian Government Research Training Program Scholarship. 
\medskip

\textbf{Data Availability Statement} \par
The source code and datasets generated and / or analysed during the current study are available from the corresponding author upon reasonable request.

\medskip

\renewcommand{\href}[2]{#2}
\bibliography{references}

\begin{thebibliography}{47}%
\makeatletter
\providecommand \@ifxundefined [1]{%
 \@ifx{#1\undefined}
}%
\providecommand \@ifnum [1]{%
 \ifnum #1\expandafter \@firstoftwo
 \else \expandafter \@secondoftwo
 \fi
}%
\providecommand \@ifx [1]{%
 \ifx #1\expandafter \@firstoftwo
 \else \expandafter \@secondoftwo
 \fi
}%
\providecommand \natexlab [1]{#1}%
\providecommand \enquote  [1]{``#1''}%
\providecommand \bibnamefont  [1]{#1}%
\providecommand \bibfnamefont [1]{#1}%
\providecommand \citenamefont [1]{#1}%
\providecommand \href@noop [0]{\@secondoftwo}%
\providecommand \href [0]{\begingroup \@sanitize@url \@href}%
\providecommand \@href[1]{\@@startlink{#1}\@@href}%
\providecommand \@@href[1]{\endgroup#1\@@endlink}%
\providecommand \@sanitize@url [0]{\catcode `\\12\catcode `\$12\catcode `\&12\catcode `\#12\catcode `\^12\catcode `\_12\catcode `\%12\relax}%
\providecommand \@@startlink[1]{}%
\providecommand \@@endlink[0]{}%
\providecommand \url  [0]{\begingroup\@sanitize@url \@url }%
\providecommand \@url [1]{\endgroup\@href {#1}{\urlprefix }}%
\providecommand \urlprefix  [0]{URL }%
\providecommand \Eprint [0]{\href }%
\providecommand \doibase [0]{https://doi.org/}%
\providecommand \selectlanguage [0]{\@gobble}%
\providecommand \bibinfo  [0]{\@secondoftwo}%
\providecommand \bibfield  [0]{\@secondoftwo}%
\providecommand \translation [1]{[#1]}%
\providecommand \BibitemOpen [0]{}%
\providecommand \bibitemStop [0]{}%
\providecommand \bibitemNoStop [0]{.\EOS\space}%
\providecommand \EOS [0]{\spacefactor3000\relax}%
\providecommand \BibitemShut  [1]{\csname bibitem#1\endcsname}%
\let\auto@bib@innerbib\@empty
\bibitem [{\citenamefont {Shor}(1994)}]{shor}%
  \BibitemOpen
  \bibfield  {author} {\bibinfo {author} {\bibfnamefont {P.}~\bibnamefont {Shor}},\ }\bibfield  {title} {\bibinfo {title} {Algorithms for quantum computation: discrete logarithms and factoring},\ }in\ \href {https://doi.org/10.1109/SFCS.1994.365700} {\emph {\bibinfo {booktitle} {Proceedings 35th Annual Symposium on Foundations of Computer Science}}}\ (\bibinfo  {publisher} {IEEE Xplore, Santa Fe, New Mexico},\ \bibinfo {year} {1994})\ pp.\ \bibinfo {pages} {124--134}\BibitemShut {NoStop}%
\bibitem [{\citenamefont {Grover}(1996)}]{grover}%
  \BibitemOpen
  \bibfield  {author} {\bibinfo {author} {\bibfnamefont {L.~K.}\ \bibnamefont {Grover}},\ }\bibfield  {title} {\bibinfo {title} {A fast quantum mechanical algorithm for database search},\ }in\ \href@noop {} {\emph {\bibinfo {booktitle} {Proceedings of the twenty-eighth annual ACM symposium on Theory of computing}}}\ (\bibinfo  {publisher} {Association for Computing Machinery, Philadelphia, Pennsylvania USA},\ \bibinfo {year} {1996})\ pp.\ \bibinfo {pages} {212--219}\BibitemShut {NoStop}%
\bibitem [{\citenamefont {Peruzzo}\ \emph {et~al.}(2014)\citenamefont {Peruzzo}, \citenamefont {McClean}, \citenamefont {Shadbolt}, \citenamefont {Yung}, \citenamefont {Zhou}, \citenamefont {Love}, \citenamefont {Aspuru-Guzik},\ and\ \citenamefont {O’brien}}]{VQE}%
  \BibitemOpen
  \bibfield  {author} {\bibinfo {author} {\bibfnamefont {A.}~\bibnamefont {Peruzzo}}, \bibinfo {author} {\bibfnamefont {J.}~\bibnamefont {McClean}}, \bibinfo {author} {\bibfnamefont {P.}~\bibnamefont {Shadbolt}}, \bibinfo {author} {\bibfnamefont {M.-H.}\ \bibnamefont {Yung}}, \bibinfo {author} {\bibfnamefont {X.-Q.}\ \bibnamefont {Zhou}}, \bibinfo {author} {\bibfnamefont {P.~J.}\ \bibnamefont {Love}}, \bibinfo {author} {\bibfnamefont {A.}~\bibnamefont {Aspuru-Guzik}},\ and\ \bibinfo {author} {\bibfnamefont {J.~L.}\ \bibnamefont {O’brien}},\ }\bibfield  {title} {\bibinfo {title} {A variational eigenvalue solver on a photonic quantum processor},\ }\href@noop {} {\bibfield  {journal} {\bibinfo  {journal} {Nature Communications}\ }\textbf {\bibinfo {volume} {5}},\ \bibinfo {pages} {4213} (\bibinfo {year} {2014})}\BibitemShut {NoStop}%
\bibitem [{\citenamefont {Farhi}\ \emph {et~al.}(2014)\citenamefont {Farhi}, \citenamefont {Goldstone},\ and\ \citenamefont {Gutmann}}]{QAOA}%
  \BibitemOpen
  \bibfield  {author} {\bibinfo {author} {\bibfnamefont {E.}~\bibnamefont {Farhi}}, \bibinfo {author} {\bibfnamefont {J.}~\bibnamefont {Goldstone}},\ and\ \bibinfo {author} {\bibfnamefont {S.}~\bibnamefont {Gutmann}},\ }\bibfield  {title} {\bibinfo {title} {A quantum approximate optimization algorithm},\ }\href@noop {} {\bibfield  {journal} {\bibinfo  {journal} {arXiv preprint arXiv:1411.4028}\ } (\bibinfo {year} {2014})}\BibitemShut {NoStop}%
\bibitem [{\citenamefont {Biamonte}\ \emph {et~al.}(2017)\citenamefont {Biamonte}, \citenamefont {Wittek}, \citenamefont {Pancotti}, \citenamefont {Rebentrost}, \citenamefont {Wiebe},\ and\ \citenamefont {Lloyd}}]{QML}%
  \BibitemOpen
  \bibfield  {author} {\bibinfo {author} {\bibfnamefont {J.}~\bibnamefont {Biamonte}}, \bibinfo {author} {\bibfnamefont {P.}~\bibnamefont {Wittek}}, \bibinfo {author} {\bibfnamefont {N.}~\bibnamefont {Pancotti}}, \bibinfo {author} {\bibfnamefont {P.}~\bibnamefont {Rebentrost}}, \bibinfo {author} {\bibfnamefont {N.}~\bibnamefont {Wiebe}},\ and\ \bibinfo {author} {\bibfnamefont {S.}~\bibnamefont {Lloyd}},\ }\bibfield  {title} {\bibinfo {title} {Quantum machine learning},\ }\href@noop {} {\bibfield  {journal} {\bibinfo  {journal} {Nature}\ }\textbf {\bibinfo {volume} {549}},\ \bibinfo {pages} {195} (\bibinfo {year} {2017})}\BibitemShut {NoStop}%
\bibitem [{\citenamefont {Schuld}\ \emph {et~al.}(2014)\citenamefont {Schuld}, \citenamefont {Sinayskiy},\ and\ \citenamefont {Petruccione}}]{QML_2}%
  \BibitemOpen
  \bibfield  {author} {\bibinfo {author} {\bibfnamefont {M.}~\bibnamefont {Schuld}}, \bibinfo {author} {\bibfnamefont {I.}~\bibnamefont {Sinayskiy}},\ and\ \bibinfo {author} {\bibfnamefont {F.}~\bibnamefont {Petruccione}},\ }\bibfield  {title} {\bibinfo {title} {An introduction to quantum machine learning},\ }\href {https://doi.org/10.1080/00107514.2014.964942} {\bibfield  {journal} {\bibinfo  {journal} {Contemporary Physics}\ }\textbf {\bibinfo {volume} {56}},\ \bibinfo {pages} {172} (\bibinfo {year} {2014})}\BibitemShut {NoStop}%
\bibitem [{\citenamefont {McArdle}\ \emph {et~al.}(2020)\citenamefont {McArdle}, \citenamefont {Endo}, \citenamefont {Aspuru-Guzik}, \citenamefont {Benjamin},\ and\ \citenamefont {Yuan}}]{quantum_computational_chemistry}%
  \BibitemOpen
  \bibfield  {author} {\bibinfo {author} {\bibfnamefont {S.}~\bibnamefont {McArdle}}, \bibinfo {author} {\bibfnamefont {S.}~\bibnamefont {Endo}}, \bibinfo {author} {\bibfnamefont {A.}~\bibnamefont {Aspuru-Guzik}}, \bibinfo {author} {\bibfnamefont {S.~C.}\ \bibnamefont {Benjamin}},\ and\ \bibinfo {author} {\bibfnamefont {X.}~\bibnamefont {Yuan}},\ }\bibfield  {title} {\bibinfo {title} {Quantum computational chemistry},\ }\href@noop {} {\bibfield  {journal} {\bibinfo  {journal} {Reviews of Modern Physics}\ }\textbf {\bibinfo {volume} {92}},\ \bibinfo {pages} {015003} (\bibinfo {year} {2020})}\BibitemShut {NoStop}%
\bibitem [{\citenamefont {Hollenberg}(2000)}]{quantum_biology}%
  \BibitemOpen
  \bibfield  {author} {\bibinfo {author} {\bibfnamefont {L.~C.~L.}\ \bibnamefont {Hollenberg}},\ }\bibfield  {title} {\bibinfo {title} {Fast quantum search algorithms in protein sequence comparisons:{\hspace{0.6em} }quantum bioinformatics},\ }\href {https://doi.org/10.1103/physreve.62.7532} {\bibfield  {journal} {\bibinfo  {journal} {Physical Review E}\ }\textbf {\bibinfo {volume} {62}},\ \bibinfo {pages} {7532} (\bibinfo {year} {2000})}\BibitemShut {NoStop}%
\bibitem [{\citenamefont {Or{\'{u}}s}\ \emph {et~al.}(2019)\citenamefont {Or{\'{u}}s}, \citenamefont {Mugel},\ and\ \citenamefont {Lizaso}}]{quantum_finance}%
  \BibitemOpen
  \bibfield  {author} {\bibinfo {author} {\bibfnamefont {R.}~\bibnamefont {Or{\'{u}}s}}, \bibinfo {author} {\bibfnamefont {S.}~\bibnamefont {Mugel}},\ and\ \bibinfo {author} {\bibfnamefont {E.}~\bibnamefont {Lizaso}},\ }\bibfield  {title} {\bibinfo {title} {Quantum computing for finance: Overview and prospects},\ }\href {https://doi.org/10.1016/j.revip.2019.100028} {\bibfield  {journal} {\bibinfo  {journal} {Reviews in Physics}\ }\textbf {\bibinfo {volume} {4}},\ \bibinfo {pages} {100028} (\bibinfo {year} {2019})}\BibitemShut {NoStop}%
\bibitem [{\citenamefont {Vidal}(2003)}]{classical_simulation_1}%
  \BibitemOpen
  \bibfield  {author} {\bibinfo {author} {\bibfnamefont {G.}~\bibnamefont {Vidal}},\ }\bibfield  {title} {\bibinfo {title} {Efficient classical simulation of slightly entangled quantum computations},\ }\href@noop {} {\bibfield  {journal} {\bibinfo  {journal} {Phys. Rev. Lett.}\ }\textbf {\bibinfo {volume} {91}},\ \bibinfo {pages} {147902} (\bibinfo {year} {2003})}\BibitemShut {NoStop}%
\bibitem [{\citenamefont {Verstraete}\ and\ \citenamefont {Cirac}(2004)}]{classical_simulation_2}%
  \BibitemOpen
  \bibfield  {author} {\bibinfo {author} {\bibfnamefont {F.}~\bibnamefont {Verstraete}}\ and\ \bibinfo {author} {\bibfnamefont {J.~I.}\ \bibnamefont {Cirac}},\ }\bibfield  {title} {\bibinfo {title} {Renormalization algorithms for quantum-many body systems in two and higher dimensions},\ }\href@noop {} {\bibfield  {journal} {\bibinfo  {journal} {arXiv}\ } (\bibinfo {year} {2004})},\ \Eprint {https://arxiv.org/abs/cond-mat/0407066} {cond-mat/0407066 [cond-mat.str-el]} \BibitemShut {NoStop}%
\bibitem [{\citenamefont {Verstraete}\ \emph {et~al.}(2008)\citenamefont {Verstraete}, \citenamefont {Murg},\ and\ \citenamefont {Cirac}}]{classical_simulation_3}%
  \BibitemOpen
  \bibfield  {author} {\bibinfo {author} {\bibfnamefont {F.}~\bibnamefont {Verstraete}}, \bibinfo {author} {\bibfnamefont {V.}~\bibnamefont {Murg}},\ and\ \bibinfo {author} {\bibfnamefont {J.~I.}\ \bibnamefont {Cirac}},\ }\bibfield  {title} {\bibinfo {title} {Matrix product states, projected entangled pair states, and variational renormalization group methods for quantum spin systems},\ }\href@noop {} {\bibfield  {journal} {\bibinfo  {journal} {Advances in physics}\ }\textbf {\bibinfo {volume} {57}},\ \bibinfo {pages} {143} (\bibinfo {year} {2008})}\BibitemShut {NoStop}%
\bibitem [{\citenamefont {Mooney}\ \emph {et~al.}(2019)\citenamefont {Mooney}, \citenamefont {Hill},\ and\ \citenamefont {Hollenberg}}]{bipartite_20}%
  \BibitemOpen
  \bibfield  {author} {\bibinfo {author} {\bibfnamefont {G.~J.}\ \bibnamefont {Mooney}}, \bibinfo {author} {\bibfnamefont {C.~D.}\ \bibnamefont {Hill}},\ and\ \bibinfo {author} {\bibfnamefont {L.~C.~L.}\ \bibnamefont {Hollenberg}},\ }\bibfield  {title} {\bibinfo {title} {Entanglement in a 20-qubit superconducting quantum computer},\ }\bibfield  {journal} {\bibinfo  {journal} {Scientific Reports}\ }\textbf {\bibinfo {volume} {9}},\ \href {https://doi.org/10.1038/s41598-019-49805-7} {10.1038/s41598-019-49805-7} (\bibinfo {year} {2019})\BibitemShut {NoStop}%
\bibitem [{\citenamefont {Mooney}\ \emph {et~al.}(2021{\natexlab{a}})\citenamefont {Mooney}, \citenamefont {White}, \citenamefont {Hill},\ and\ \citenamefont {Hollenberg}}]{bipartite_65}%
  \BibitemOpen
  \bibfield  {author} {\bibinfo {author} {\bibfnamefont {G.~J.}\ \bibnamefont {Mooney}}, \bibinfo {author} {\bibfnamefont {G.~A.~L.}\ \bibnamefont {White}}, \bibinfo {author} {\bibfnamefont {C.~D.}\ \bibnamefont {Hill}},\ and\ \bibinfo {author} {\bibfnamefont {L.~C.~L.}\ \bibnamefont {Hollenberg}},\ }\bibfield  {title} {\bibinfo {title} {Whole-device entanglement in a 65-qubit superconducting quantum computer},\ }\href {https://doi.org/10.1002/qute.202100061} {\bibfield  {journal} {\bibinfo  {journal} {Advanced Quantum Technologies}\ }\textbf {\bibinfo {volume} {4}},\ \bibinfo {pages} {2100061} (\bibinfo {year} {2021}{\natexlab{a}})}\BibitemShut {NoStop}%
\bibitem [{\citenamefont {Moses~et al.}(2023)}]{GHZ_32}%
  \BibitemOpen
  \bibfield  {author} {\bibinfo {author} {\bibfnamefont {S.~A.}\ \bibnamefont {Moses~et al.}},\ }\bibfield  {title} {\bibinfo {title} {A race-track trapped-ion quantum processor},\ }\href {https://doi.org/10.1103/PhysRevX.13.041052} {\bibfield  {journal} {\bibinfo  {journal} {Phys. Rev. X}\ }\textbf {\bibinfo {volume} {13}},\ \bibinfo {pages} {041052} (\bibinfo {year} {2023})}\BibitemShut {NoStop}%
\bibitem [{\citenamefont {Kam}\ \emph {et~al.}(2024)\citenamefont {Kam}, \citenamefont {Kang}, \citenamefont {Hill}, \citenamefont {Mooney},\ and\ \citenamefont {Hollenberg}}]{fidel_paper}%
  \BibitemOpen
  \bibfield  {author} {\bibinfo {author} {\bibfnamefont {J.~F.}\ \bibnamefont {Kam}}, \bibinfo {author} {\bibfnamefont {H.}~\bibnamefont {Kang}}, \bibinfo {author} {\bibfnamefont {C.~D.}\ \bibnamefont {Hill}}, \bibinfo {author} {\bibfnamefont {G.~J.}\ \bibnamefont {Mooney}},\ and\ \bibinfo {author} {\bibfnamefont {L.~C.}\ \bibnamefont {Hollenberg}},\ }\bibfield  {title} {\bibinfo {title} {Characterization of entanglement on superconducting quantum computers of up to 414 qubits},\ }\href@noop {} {\bibfield  {journal} {\bibinfo  {journal} {Physical Review Research}\ }\textbf {\bibinfo {volume} {6}},\ \bibinfo {pages} {033155} (\bibinfo {year} {2024})}\BibitemShut {NoStop}%
\bibitem [{\citenamefont {Cao}\ \emph {et~al.}(2023)\citenamefont {Cao}, \citenamefont {Wu}, \citenamefont {Chen}, \citenamefont {Gong}, \citenamefont {Wu}, \citenamefont {Ye}, \citenamefont {Zha}, \citenamefont {Qian}, \citenamefont {Ying}, \citenamefont {Guo} \emph {et~al.}}]{cao51entanglement}%
  \BibitemOpen
  \bibfield  {author} {\bibinfo {author} {\bibfnamefont {S.}~\bibnamefont {Cao}}, \bibinfo {author} {\bibfnamefont {B.}~\bibnamefont {Wu}}, \bibinfo {author} {\bibfnamefont {F.}~\bibnamefont {Chen}}, \bibinfo {author} {\bibfnamefont {M.}~\bibnamefont {Gong}}, \bibinfo {author} {\bibfnamefont {Y.}~\bibnamefont {Wu}}, \bibinfo {author} {\bibfnamefont {Y.}~\bibnamefont {Ye}}, \bibinfo {author} {\bibfnamefont {C.}~\bibnamefont {Zha}}, \bibinfo {author} {\bibfnamefont {H.}~\bibnamefont {Qian}}, \bibinfo {author} {\bibfnamefont {C.}~\bibnamefont {Ying}}, \bibinfo {author} {\bibfnamefont {S.}~\bibnamefont {Guo}}, \emph {et~al.},\ }\bibfield  {title} {\bibinfo {title} {Generation of genuine entanglement up to 51 superconducting qubits},\ }\href@noop {} {\bibfield  {journal} {\bibinfo  {journal} {Nature}\ ,\ \bibinfo {pages} {1}} (\bibinfo {year} {2023})}\BibitemShut {NoStop}%
\bibitem [{\citenamefont {Bennett}\ \emph {et~al.}(1993)\citenamefont {Bennett}, \citenamefont {Brassard}, \citenamefont {Cr\'epeau}, \citenamefont {Jozsa}, \citenamefont {Peres},\ and\ \citenamefont {Wootters}}]{Teleportation}%
  \BibitemOpen
  \bibfield  {author} {\bibinfo {author} {\bibfnamefont {C.~H.}\ \bibnamefont {Bennett}}, \bibinfo {author} {\bibfnamefont {G.}~\bibnamefont {Brassard}}, \bibinfo {author} {\bibfnamefont {C.}~\bibnamefont {Cr\'epeau}}, \bibinfo {author} {\bibfnamefont {R.}~\bibnamefont {Jozsa}}, \bibinfo {author} {\bibfnamefont {A.}~\bibnamefont {Peres}},\ and\ \bibinfo {author} {\bibfnamefont {W.~K.}\ \bibnamefont {Wootters}},\ }\bibfield  {title} {\bibinfo {title} {Teleporting an unknown quantum state via dual classical and {Einstein}-{Podolsky}-{Rosen} channels},\ }\href {https://doi.org/10.1103/PhysRevLett.70.1895} {\bibfield  {journal} {\bibinfo  {journal} {Phys. Rev. Lett.}\ }\textbf {\bibinfo {volume} {70}},\ \bibinfo {pages} {1895} (\bibinfo {year} {1993})}\BibitemShut {NoStop}%
\bibitem [{\citenamefont {Bouwmeester}\ \emph {et~al.}(1997)\citenamefont {Bouwmeester}, \citenamefont {Pan}, \citenamefont {Mattle}, \citenamefont {Eibl}, \citenamefont {Weinfurter},\ and\ \citenamefont {Zeilinger}}]{first_experimental_teleportation}%
  \BibitemOpen
  \bibfield  {author} {\bibinfo {author} {\bibfnamefont {D.}~\bibnamefont {Bouwmeester}}, \bibinfo {author} {\bibfnamefont {J.-W.}\ \bibnamefont {Pan}}, \bibinfo {author} {\bibfnamefont {K.}~\bibnamefont {Mattle}}, \bibinfo {author} {\bibfnamefont {M.}~\bibnamefont {Eibl}}, \bibinfo {author} {\bibfnamefont {H.}~\bibnamefont {Weinfurter}},\ and\ \bibinfo {author} {\bibfnamefont {A.}~\bibnamefont {Zeilinger}},\ }\bibfield  {title} {\bibinfo {title} {Experimental quantum teleportation},\ }\href@noop {} {\bibfield  {journal} {\bibinfo  {journal} {Nature}\ }\textbf {\bibinfo {volume} {390}},\ \bibinfo {pages} {575} (\bibinfo {year} {1997})}\BibitemShut {NoStop}%
\bibitem [{\citenamefont {Kim}\ \emph {et~al.}(2001)\citenamefont {Kim}, \citenamefont {Kulik},\ and\ \citenamefont {Shih}}]{teleportation_of_polarization_state}%
  \BibitemOpen
  \bibfield  {author} {\bibinfo {author} {\bibfnamefont {Y.-H.}\ \bibnamefont {Kim}}, \bibinfo {author} {\bibfnamefont {S.~P.}\ \bibnamefont {Kulik}},\ and\ \bibinfo {author} {\bibfnamefont {Y.}~\bibnamefont {Shih}},\ }\bibfield  {title} {\bibinfo {title} {Quantum teleportation of a polarization state with a complete {Bell} state measurement},\ }\href {https://doi.org/10.1103/PhysRevLett.86.1370} {\bibfield  {journal} {\bibinfo  {journal} {Phys. Rev. Lett.}\ }\textbf {\bibinfo {volume} {86}},\ \bibinfo {pages} {1370} (\bibinfo {year} {2001})}\BibitemShut {NoStop}%
\bibitem [{\citenamefont {Bennett}\ and\ \citenamefont {Wiesner}(1992)}]{dense_coding}%
  \BibitemOpen
  \bibfield  {author} {\bibinfo {author} {\bibfnamefont {C.~H.}\ \bibnamefont {Bennett}}\ and\ \bibinfo {author} {\bibfnamefont {S.~J.}\ \bibnamefont {Wiesner}},\ }\bibfield  {title} {\bibinfo {title} {Communication via one-and two-particle operators on {Einstein}-{Podolsky}-{Rosen} states},\ }\href@noop {} {\bibfield  {journal} {\bibinfo  {journal} {Physical Review Letters}\ }\textbf {\bibinfo {volume} {69}},\ \bibinfo {pages} {2881} (\bibinfo {year} {1992})}\BibitemShut {NoStop}%
\bibitem [{\citenamefont {Wang}\ \emph {et~al.}(2014{\natexlab{a}})\citenamefont {Wang}, \citenamefont {Yu}, \citenamefont {Lu},\ and\ \citenamefont {Gong}}]{quantum_communication}%
  \BibitemOpen
  \bibfield  {author} {\bibinfo {author} {\bibfnamefont {K.}~\bibnamefont {Wang}}, \bibinfo {author} {\bibfnamefont {X.-T.}\ \bibnamefont {Yu}}, \bibinfo {author} {\bibfnamefont {S.-L.}\ \bibnamefont {Lu}},\ and\ \bibinfo {author} {\bibfnamefont {Y.-X.}\ \bibnamefont {Gong}},\ }\bibfield  {title} {\bibinfo {title} {Quantum wireless multihop communication based on arbitrary {Bell} pairs and teleportation},\ }\href {https://doi.org/10.1103/PhysRevA.89.022329} {\bibfield  {journal} {\bibinfo  {journal} {Phys. Rev. A}\ }\textbf {\bibinfo {volume} {89}},\ \bibinfo {pages} {022329} (\bibinfo {year} {2014}{\natexlab{a}})}\BibitemShut {NoStop}%
\bibitem [{\citenamefont {Yokoyama}\ \emph {et~al.}(2013)\citenamefont {Yokoyama}, \citenamefont {Ukai}, \citenamefont {Armstrong}, \citenamefont {Sornphiphatphong}, \citenamefont {Kaji}, \citenamefont {Suzuki}, \citenamefont {Yoshikawa}, \citenamefont {Yonezawa}, \citenamefont {Menicucci},\ and\ \citenamefont {Furusawa}}]{detect_entanglement_via_teleportation}%
  \BibitemOpen
  \bibfield  {author} {\bibinfo {author} {\bibfnamefont {S.}~\bibnamefont {Yokoyama}}, \bibinfo {author} {\bibfnamefont {R.}~\bibnamefont {Ukai}}, \bibinfo {author} {\bibfnamefont {S.~C.}\ \bibnamefont {Armstrong}}, \bibinfo {author} {\bibfnamefont {C.}~\bibnamefont {Sornphiphatphong}}, \bibinfo {author} {\bibfnamefont {T.}~\bibnamefont {Kaji}}, \bibinfo {author} {\bibfnamefont {S.}~\bibnamefont {Suzuki}}, \bibinfo {author} {\bibfnamefont {J.-i.}\ \bibnamefont {Yoshikawa}}, \bibinfo {author} {\bibfnamefont {H.}~\bibnamefont {Yonezawa}}, \bibinfo {author} {\bibfnamefont {N.~C.}\ \bibnamefont {Menicucci}},\ and\ \bibinfo {author} {\bibfnamefont {A.}~\bibnamefont {Furusawa}},\ }\bibfield  {title} {\bibinfo {title} {Ultra-large-scale continuous-variable cluster states multiplexed in the time domain},\ }\href@noop {} {\bibfield  {journal} {\bibinfo  {journal} {Nature Photonics}\ }\textbf {\bibinfo {volume} {7}},\ \bibinfo {pages} {982} (\bibinfo {year} {2013})}\BibitemShut {NoStop}%
\bibitem [{\citenamefont {Olmschenk}\ \emph {et~al.}(2009)\citenamefont {Olmschenk}, \citenamefont {Matsukevich}, \citenamefont {Maunz}, \citenamefont {Hayes}, \citenamefont {Duan},\ and\ \citenamefont {Monroe}}]{teleportation_on_distant_qubits}%
  \BibitemOpen
  \bibfield  {author} {\bibinfo {author} {\bibfnamefont {S.}~\bibnamefont {Olmschenk}}, \bibinfo {author} {\bibfnamefont {D.}~\bibnamefont {Matsukevich}}, \bibinfo {author} {\bibfnamefont {P.}~\bibnamefont {Maunz}}, \bibinfo {author} {\bibfnamefont {D.}~\bibnamefont {Hayes}}, \bibinfo {author} {\bibfnamefont {L.-M.}\ \bibnamefont {Duan}},\ and\ \bibinfo {author} {\bibfnamefont {C.}~\bibnamefont {Monroe}},\ }\bibfield  {title} {\bibinfo {title} {Quantum teleportation between distant matter qubits},\ }\href@noop {} {\bibfield  {journal} {\bibinfo  {journal} {Science}\ }\textbf {\bibinfo {volume} {323}},\ \bibinfo {pages} {486} (\bibinfo {year} {2009})}\BibitemShut {NoStop}%
\bibitem [{\citenamefont {Hermans}\ \emph {et~al.}(2022)\citenamefont {Hermans}, \citenamefont {Pompili}, \citenamefont {Beukers}, \citenamefont {Baier}, \citenamefont {Borregaard},\ and\ \citenamefont {Hanson}}]{teleportation_on_distant_qubits2}%
  \BibitemOpen
  \bibfield  {author} {\bibinfo {author} {\bibfnamefont {S.}~\bibnamefont {Hermans}}, \bibinfo {author} {\bibfnamefont {M.}~\bibnamefont {Pompili}}, \bibinfo {author} {\bibfnamefont {H.}~\bibnamefont {Beukers}}, \bibinfo {author} {\bibfnamefont {S.}~\bibnamefont {Baier}}, \bibinfo {author} {\bibfnamefont {J.}~\bibnamefont {Borregaard}},\ and\ \bibinfo {author} {\bibfnamefont {R.}~\bibnamefont {Hanson}},\ }\bibfield  {title} {\bibinfo {title} {Qubit teleportation between non-neighbouring nodes in a quantum network},\ }\href@noop {} {\bibfield  {journal} {\bibinfo  {journal} {Nature}\ }\textbf {\bibinfo {volume} {605}},\ \bibinfo {pages} {663} (\bibinfo {year} {2022})}\BibitemShut {NoStop}%
\bibitem [{\citenamefont {Beverland}\ \emph {et~al.}(2022)\citenamefont {Beverland}, \citenamefont {Kliuchnikov},\ and\ \citenamefont {Schoute}}]{long_range_gates_2}%
  \BibitemOpen
  \bibfield  {author} {\bibinfo {author} {\bibfnamefont {M.}~\bibnamefont {Beverland}}, \bibinfo {author} {\bibfnamefont {V.}~\bibnamefont {Kliuchnikov}},\ and\ \bibinfo {author} {\bibfnamefont {E.}~\bibnamefont {Schoute}},\ }\bibfield  {title} {\bibinfo {title} {Surface code compilation via edge-disjoint paths},\ }\href {https://doi.org/10.1103/PRXQuantum.3.020342} {\bibfield  {journal} {\bibinfo  {journal} {PRX Quantum}\ }\textbf {\bibinfo {volume} {3}},\ \bibinfo {pages} {020342} (\bibinfo {year} {2022})}\BibitemShut {NoStop}%
\bibitem [{\citenamefont {B{\"a}umer}\ \emph {et~al.}(2024{\natexlab{a}})\citenamefont {B{\"a}umer}, \citenamefont {Tripathi}, \citenamefont {Wang}, \citenamefont {Rall}, \citenamefont {Chen}, \citenamefont {Majumder}, \citenamefont {Seif},\ and\ \citenamefont {Minev}}]{long_range_gates}%
  \BibitemOpen
  \bibfield  {author} {\bibinfo {author} {\bibfnamefont {E.}~\bibnamefont {B{\"a}umer}}, \bibinfo {author} {\bibfnamefont {V.}~\bibnamefont {Tripathi}}, \bibinfo {author} {\bibfnamefont {D.~S.}\ \bibnamefont {Wang}}, \bibinfo {author} {\bibfnamefont {P.}~\bibnamefont {Rall}}, \bibinfo {author} {\bibfnamefont {E.~H.}\ \bibnamefont {Chen}}, \bibinfo {author} {\bibfnamefont {S.}~\bibnamefont {Majumder}}, \bibinfo {author} {\bibfnamefont {A.}~\bibnamefont {Seif}},\ and\ \bibinfo {author} {\bibfnamefont {Z.~K.}\ \bibnamefont {Minev}},\ }\bibfield  {title} {\bibinfo {title} {Efficient long-range entanglement using dynamic circuits},\ }\href@noop {} {\bibfield  {journal} {\bibinfo  {journal} {PRX Quantum}\ }\textbf {\bibinfo {volume} {5}},\ \bibinfo {pages} {030339} (\bibinfo {year} {2024}{\natexlab{a}})}\BibitemShut {NoStop}%
\bibitem [{\citenamefont {Bose}\ \emph {et~al.}(1998)\citenamefont {Bose}, \citenamefont {Vedral},\ and\ \citenamefont {Knight}}]{entanglement_swapping}%
  \BibitemOpen
  \bibfield  {author} {\bibinfo {author} {\bibfnamefont {S.}~\bibnamefont {Bose}}, \bibinfo {author} {\bibfnamefont {V.}~\bibnamefont {Vedral}},\ and\ \bibinfo {author} {\bibfnamefont {P.~L.}\ \bibnamefont {Knight}},\ }\bibfield  {title} {\bibinfo {title} {Multiparticle generalization of entanglement swapping},\ }\href {https://doi.org/10.1103/PhysRevA.57.822} {\bibfield  {journal} {\bibinfo  {journal} {Phys. Rev. A}\ }\textbf {\bibinfo {volume} {57}},\ \bibinfo {pages} {822} (\bibinfo {year} {1998})}\BibitemShut {NoStop}%
\bibitem [{\citenamefont {Jozsa}(2006)}]{dynamic_circuit}%
  \BibitemOpen
  \bibfield  {author} {\bibinfo {author} {\bibfnamefont {R.}~\bibnamefont {Jozsa}},\ }\bibfield  {title} {\bibinfo {title} {An introduction to measurement based quantum computation},\ }\href@noop {} {\bibfield  {journal} {\bibinfo  {journal} {NATO Science Series, III: Computer and Systems Sciences. Quantum Information Processing-From Theory to Experiment}\ }\textbf {\bibinfo {volume} {199}},\ \bibinfo {pages} {137} (\bibinfo {year} {2006})}\BibitemShut {NoStop}%
\bibitem [{\citenamefont {Koch}\ \emph {et~al.}(2007)\citenamefont {Koch}, \citenamefont {Yu}, \citenamefont {Gambetta}, \citenamefont {Houck}, \citenamefont {Schuster}, \citenamefont {Majer}, \citenamefont {Blais}, \citenamefont {Devoret}, \citenamefont {Girvin},\ and\ \citenamefont {Schoelkopf}}]{transmon_qubits}%
  \BibitemOpen
  \bibfield  {author} {\bibinfo {author} {\bibfnamefont {J.}~\bibnamefont {Koch}}, \bibinfo {author} {\bibfnamefont {T.~M.}\ \bibnamefont {Yu}}, \bibinfo {author} {\bibfnamefont {J.}~\bibnamefont {Gambetta}}, \bibinfo {author} {\bibfnamefont {A.~A.}\ \bibnamefont {Houck}}, \bibinfo {author} {\bibfnamefont {D.~I.}\ \bibnamefont {Schuster}}, \bibinfo {author} {\bibfnamefont {J.}~\bibnamefont {Majer}}, \bibinfo {author} {\bibfnamefont {A.}~\bibnamefont {Blais}}, \bibinfo {author} {\bibfnamefont {M.~H.}\ \bibnamefont {Devoret}}, \bibinfo {author} {\bibfnamefont {S.~M.}\ \bibnamefont {Girvin}},\ and\ \bibinfo {author} {\bibfnamefont {R.~J.}\ \bibnamefont {Schoelkopf}},\ }\bibfield  {title} {\bibinfo {title} {Charge-insensitive qubit design derived from the {Cooper} pair box},\ }\href {https://doi.org/10.1103/PhysRevA.76.042319} {\bibfield  {journal} {\bibinfo  {journal} {Phys. Rev. A}\ }\textbf {\bibinfo {volume} {76}},\ \bibinfo {pages} {042319} (\bibinfo {year} {2007})}\BibitemShut {NoStop}%
\bibitem [{\citenamefont {Wang}\ \emph {et~al.}(2014{\natexlab{b}})\citenamefont {Wang}, \citenamefont {Yu}, \citenamefont {Lu},\ and\ \citenamefont {Gong}}]{discriminant_vec}%
  \BibitemOpen
  \bibfield  {author} {\bibinfo {author} {\bibfnamefont {K.}~\bibnamefont {Wang}}, \bibinfo {author} {\bibfnamefont {X.-T.}\ \bibnamefont {Yu}}, \bibinfo {author} {\bibfnamefont {S.-L.}\ \bibnamefont {Lu}},\ and\ \bibinfo {author} {\bibfnamefont {Y.-X.}\ \bibnamefont {Gong}},\ }\bibfield  {title} {\bibinfo {title} {Quantum wireless multihop communication based on arbitrary {Bell} pairs and teleportation},\ }\href {https://doi.org/10.1103/PhysRevA.89.022329} {\bibfield  {journal} {\bibinfo  {journal} {Phys. Rev. A}\ }\textbf {\bibinfo {volume} {89}},\ \bibinfo {pages} {022329} (\bibinfo {year} {2014}{\natexlab{b}})}\BibitemShut {NoStop}%
\bibitem [{\citenamefont {Vidal}\ and\ \citenamefont {Werner}(2002)}]{Negativity}%
  \BibitemOpen
  \bibfield  {author} {\bibinfo {author} {\bibfnamefont {G.}~\bibnamefont {Vidal}}\ and\ \bibinfo {author} {\bibfnamefont {R.~F.}\ \bibnamefont {Werner}},\ }\bibfield  {title} {\bibinfo {title} {Computable measure of entanglement},\ }\href {https://doi.org/10.1103/PhysRevA.65.032314} {\bibfield  {journal} {\bibinfo  {journal} {Phys. Rev. A}\ }\textbf {\bibinfo {volume} {65}},\ \bibinfo {pages} {032314} (\bibinfo {year} {2002})}\BibitemShut {NoStop}%
\bibitem [{\citenamefont {Shapourian}\ \emph {et~al.}(2017)\citenamefont {Shapourian}, \citenamefont {Shiozaki},\ and\ \citenamefont {Ryu}}]{partial_transpose}%
  \BibitemOpen
  \bibfield  {author} {\bibinfo {author} {\bibfnamefont {H.}~\bibnamefont {Shapourian}}, \bibinfo {author} {\bibfnamefont {K.}~\bibnamefont {Shiozaki}},\ and\ \bibinfo {author} {\bibfnamefont {S.}~\bibnamefont {Ryu}},\ }\bibfield  {title} {\bibinfo {title} {Partial time-reversal transformation and entanglement negativity in fermionic systems},\ }\href@noop {} {\bibfield  {journal} {\bibinfo  {journal} {Physical Review B}\ }\textbf {\bibinfo {volume} {95}},\ \bibinfo {pages} {165101} (\bibinfo {year} {2017})}\BibitemShut {NoStop}%
\bibitem [{\citenamefont {Nielsen}\ and\ \citenamefont {Chuang}(2022)}]{nielsen_chuang_2022}%
  \BibitemOpen
  \bibfield  {author} {\bibinfo {author} {\bibfnamefont {M.~A.}\ \bibnamefont {Nielsen}}\ and\ \bibinfo {author} {\bibfnamefont {I.~L.}\ \bibnamefont {Chuang}},\ }\href@noop {} {\emph {\bibinfo {title} {Quantum Computation and Quantum Information}}}\ (\bibinfo  {publisher} {Cambridge University Press},\ \bibinfo {year} {2022})\BibitemShut {NoStop}%
\bibitem [{\citenamefont {Nation}\ \emph {et~al.}(2021)\citenamefont {Nation}, \citenamefont {Kang}, \citenamefont {Sundaresan},\ and\ \citenamefont {Gambetta}}]{QREM}%
  \BibitemOpen
  \bibfield  {author} {\bibinfo {author} {\bibfnamefont {P.~D.}\ \bibnamefont {Nation}}, \bibinfo {author} {\bibfnamefont {H.}~\bibnamefont {Kang}}, \bibinfo {author} {\bibfnamefont {N.}~\bibnamefont {Sundaresan}},\ and\ \bibinfo {author} {\bibfnamefont {J.~M.}\ \bibnamefont {Gambetta}},\ }\bibfield  {title} {\bibinfo {title} {Scalable mitigation of measurement errors on quantum computers},\ }\href {https://doi.org/10.1103/PRXQuantum.2.040326} {\bibfield  {journal} {\bibinfo  {journal} {PRX Quantum}\ }\textbf {\bibinfo {volume} {2}},\ \bibinfo {pages} {040326} (\bibinfo {year} {2021})}\BibitemShut {NoStop}%
\bibitem [{\citenamefont {Mooney}\ \emph {et~al.}(2021{\natexlab{b}})\citenamefont {Mooney}, \citenamefont {White}, \citenamefont {Hill},\ and\ \citenamefont {Hollenberg}}]{Mooney_2021}%
  \BibitemOpen
  \bibfield  {author} {\bibinfo {author} {\bibfnamefont {G.~J.}\ \bibnamefont {Mooney}}, \bibinfo {author} {\bibfnamefont {G.~A.~L.}\ \bibnamefont {White}}, \bibinfo {author} {\bibfnamefont {C.~D.}\ \bibnamefont {Hill}},\ and\ \bibinfo {author} {\bibfnamefont {L.~C.~L.}\ \bibnamefont {Hollenberg}},\ }\bibfield  {title} {\bibinfo {title} {Generation and verification of 27-qubit {Greenberger}-{Horne}-{Zeilinger} states in a superconducting quantum computer},\ }\href {https://doi.org/10.1088/2399-6528/ac1df7} {\bibfield  {journal} {\bibinfo  {journal} {Journal of Physics Communications}\ }\textbf {\bibinfo {volume} {5}},\ \bibinfo {pages} {095004} (\bibinfo {year} {2021}{\natexlab{b}})}\BibitemShut {NoStop}%
\bibitem [{\citenamefont {Yang}\ \emph {et~al.}(2022)\citenamefont {Yang}, \citenamefont {Raymond},\ and\ \citenamefont {Uno}}]{bo_yang_qrem}%
  \BibitemOpen
  \bibfield  {author} {\bibinfo {author} {\bibfnamefont {B.}~\bibnamefont {Yang}}, \bibinfo {author} {\bibfnamefont {R.}~\bibnamefont {Raymond}},\ and\ \bibinfo {author} {\bibfnamefont {S.}~\bibnamefont {Uno}},\ }\bibfield  {title} {\bibinfo {title} {Efficient quantum readout-error mitigation for sparse measurement outcomes of near-term quantum devices},\ }\href@noop {} {\bibfield  {journal} {\bibinfo  {journal} {Physical Review A}\ }\textbf {\bibinfo {volume} {106}},\ \bibinfo {pages} {012423} (\bibinfo {year} {2022})}\BibitemShut {NoStop}%
\bibitem [{\citenamefont {Michelot}(1986)}]{closest_pvec}%
  \BibitemOpen
  \bibfield  {author} {\bibinfo {author} {\bibfnamefont {C.}~\bibnamefont {Michelot}},\ }\bibfield  {title} {\bibinfo {title} {A finite algorithm for finding the projection of a point onto the canonical simplex of $\mathbb{R}^{n}$},\ }\href@noop {} {\bibfield  {journal} {\bibinfo  {journal} {Journal of Optimization Theory and Applications}\ }\textbf {\bibinfo {volume} {50}},\ \bibinfo {pages} {195} (\bibinfo {year} {1986})}\BibitemShut {NoStop}%
\bibitem [{\citenamefont {Smolin}\ \emph {et~al.}(2012)\citenamefont {Smolin}, \citenamefont {Gambetta},\ and\ \citenamefont {Smith}}]{closest_physical_rhos}%
  \BibitemOpen
  \bibfield  {author} {\bibinfo {author} {\bibfnamefont {J.~A.}\ \bibnamefont {Smolin}}, \bibinfo {author} {\bibfnamefont {J.~M.}\ \bibnamefont {Gambetta}},\ and\ \bibinfo {author} {\bibfnamefont {G.}~\bibnamefont {Smith}},\ }\bibfield  {title} {\bibinfo {title} {Efficient method for computing the maximum-likelihood quantum state from measurements with additive {Gaussian} noise},\ }\href {https://doi.org/10.1103/PhysRevLett.108.070502} {\bibfield  {journal} {\bibinfo  {journal} {Phys. Rev. Lett.}\ }\textbf {\bibinfo {volume} {108}},\ \bibinfo {pages} {070502} (\bibinfo {year} {2012})}\BibitemShut {NoStop}%
\bibitem [{\citenamefont {Nation}\ and\ \citenamefont {Treinish}(2023)}]{VF2++}%
  \BibitemOpen
  \bibfield  {author} {\bibinfo {author} {\bibfnamefont {P.~D.}\ \bibnamefont {Nation}}\ and\ \bibinfo {author} {\bibfnamefont {M.}~\bibnamefont {Treinish}},\ }\bibfield  {title} {\bibinfo {title} {Suppressing quantum circuit errors due to system variability},\ }\href@noop {} {\bibfield  {journal} {\bibinfo  {journal} {PRX Quantum}\ }\textbf {\bibinfo {volume} {4}},\ \bibinfo {pages} {010327} (\bibinfo {year} {2023})}\BibitemShut {NoStop}%
\bibitem [{map(2022)}]{mapomatic}%
  \BibitemOpen
  \href@noop {} {}\bibinfo {howpublished} {\url{https://github.com/qiskit-community/mapomatic}} (\bibinfo {year} {2022})\BibitemShut {NoStop}%
\bibitem [{\citenamefont {Kang}\ and\ \citenamefont {Kam}(2024)}]{Program_Codes}%
  \BibitemOpen
  \bibfield  {author} {\bibinfo {author} {\bibfnamefont {H.}~\bibnamefont {Kang}}\ and\ \bibinfo {author} {\bibfnamefont {J.~F.}\ \bibnamefont {Kam}},\ }\href@noop {} {\bibinfo {title} {teleportation\_code\_{IBM}}},\ \bibinfo {howpublished} {\url{https://github.com/KangHaiYue/teleportation_code_IBM}} (\bibinfo {year} {2024})\BibitemShut {NoStop}%
\bibitem [{Sup()}]{Sup_materials}%
  \BibitemOpen
  \href@noop {} {}\bibinfo {note} {See Supplemental Material at [URL will be inserted by publisher] for the parameters of the quantum computer \textit{ibm\_sherbrooke} at the times of execution.}\BibitemShut {Stop}%
\bibitem [{\citenamefont {Raussendorf}\ \emph {et~al.}(2003)\citenamefont {Raussendorf}, \citenamefont {Browne},\ and\ \citenamefont {Briegel}}]{one_way_qc}%
  \BibitemOpen
  \bibfield  {author} {\bibinfo {author} {\bibfnamefont {R.}~\bibnamefont {Raussendorf}}, \bibinfo {author} {\bibfnamefont {D.~E.}\ \bibnamefont {Browne}},\ and\ \bibinfo {author} {\bibfnamefont {H.~J.}\ \bibnamefont {Briegel}},\ }\bibfield  {title} {\bibinfo {title} {Measurement-based quantum computation on cluster states},\ }\bibfield  {journal} {\bibinfo  {journal} {Physical Review A}\ }\textbf {\bibinfo {volume} {68}},\ \href {https://doi.org/10.1103/physreva.68.022312} {10.1103/physreva.68.022312} (\bibinfo {year} {2003})\BibitemShut {NoStop}%
\bibitem [{\citenamefont {Uhrig}(2007)}]{dynamical_decoupling}%
  \BibitemOpen
  \bibfield  {author} {\bibinfo {author} {\bibfnamefont {G.~S.}\ \bibnamefont {Uhrig}},\ }\bibfield  {title} {\bibinfo {title} {Keeping a quantum bit alive by optimized $\pi$-pulse sequences},\ }\href@noop {} {\bibfield  {journal} {\bibinfo  {journal} {Physical Review Letters}\ }\textbf {\bibinfo {volume} {98}},\ \bibinfo {pages} {100504} (\bibinfo {year} {2007})}\BibitemShut {NoStop}%
\bibitem [{\citenamefont {B{\"a}umer}\ \emph {et~al.}(2024{\natexlab{b}})\citenamefont {B{\"a}umer}, \citenamefont {Tripathi}, \citenamefont {Seif}, \citenamefont {Lidar},\ and\ \citenamefont {Wang}}]{feed_forward_time}%
  \BibitemOpen
  \bibfield  {author} {\bibinfo {author} {\bibfnamefont {E.}~\bibnamefont {B{\"a}umer}}, \bibinfo {author} {\bibfnamefont {V.}~\bibnamefont {Tripathi}}, \bibinfo {author} {\bibfnamefont {A.}~\bibnamefont {Seif}}, \bibinfo {author} {\bibfnamefont {D.}~\bibnamefont {Lidar}},\ and\ \bibinfo {author} {\bibfnamefont {D.~S.}\ \bibnamefont {Wang}},\ }\bibfield  {title} {\bibinfo {title} {Quantum fourier transform using dynamic circuits},\ }\href@noop {} {\bibfield  {journal} {\bibinfo  {journal} {Phys. Rev. Lett.}\ }\textbf {\bibinfo {volume} {133}},\ \bibinfo {pages} {150602} (\bibinfo {year} {2024}{\natexlab{b}})}\BibitemShut {NoStop}%
\bibitem [{\citenamefont {Fowler}\ \emph {et~al.}(2012)\citenamefont {Fowler}, \citenamefont {Mariantoni}, \citenamefont {Martinis},\ and\ \citenamefont {Cleland}}]{Surface_codes}%
  \BibitemOpen
  \bibfield  {author} {\bibinfo {author} {\bibfnamefont {A.~G.}\ \bibnamefont {Fowler}}, \bibinfo {author} {\bibfnamefont {M.}~\bibnamefont {Mariantoni}}, \bibinfo {author} {\bibfnamefont {J.~M.}\ \bibnamefont {Martinis}},\ and\ \bibinfo {author} {\bibfnamefont {A.~N.}\ \bibnamefont {Cleland}},\ }\bibfield  {title} {\bibinfo {title} {Surface codes: Towards practical large-scale quantum computation},\ }\bibfield  {journal} {\bibinfo  {journal} {Physical Review A}\ }\textbf {\bibinfo {volume} {86}},\ \href {https://doi.org/10.1103/physreva.86.032324} {10.1103/physreva.86.032324} (\bibinfo {year} {2012})\BibitemShut {NoStop}%
\end{thebibliography}%

\onecolumngrid
\appendix
\section{Mathematical details of teleportation}\label{appendix}

\begin{customthm}{A1} \label{thm:teleportation across one qubit}
\textit{Let $\ket{\Phi}=CZ_{1,2}\ket{\phi}_{0,1}\otimes\ket{+}_{2}$, be a three-qubit quantum state defined on Hilbert space $\mathscr{H}_{0,1,2}$ with $\ket{\phi}_{0,1}$ as an arbitrary two-qubit quantum state rests on qubit 0 and 1 entangled with qubit 2 in the $X$-basis state $\ket{+}_{2}$ via Controlled-$Z$ gate, where $\ket{+}\coloneqq\frac{1}{\sqrt{2}}(\ket{0}+\ket{1})$ in the computational $Z$-basis.}

\textit{If qubit 1 is measured in the $X$-basis with outcome $s$, the state $\ket{\phi}_{0,1}$ will be teleported to $\ket{\phi}_{0,2}$ up to the local transformation that depends on the outcome $s$, yielding}
\begin{equation}
    {\ket{\Phi}}^{(\text{Teleported})}=\ket{s}_{1}\otimes \underbrace{X_{2}^{s_{1}}H_{2}\ket{\phi}_{0,2}}_{\ket{\psi}^{(\text{Transformed})}_{0,2}},
\end{equation}
\textit{where $s\in\{0,1\}$.}
\end{customthm}
\begin{proof}[Proof.]
In the computational $Z$-basis, we consider the state $\ket{\phi}$ to be written as
\begin{equation}
    \ket{\phi}=\alpha\ket{00}+\beta\ket{01}+\gamma\ket{10}+\delta\ket{11},
\end{equation}
where $\alpha,\beta,\gamma,\delta \in \mathbb{C}$. Hence, the state $\ket{\Phi}$ before teleportation is expressed as
\begin{equation}
\begin{aligned}
    \ket{\Phi}&=CZ_{1,2}(\alpha\ket{00}+\beta\ket{01}+\gamma\ket{10}+\delta\ket{11})_{0,1}\frac{(\ket{0}+\ket{1})_2}{\sqrt{2}}\\
    &=\frac{1}{\sqrt{2}}\bigl(\alpha(\ket{000}+\ket{001})+\beta(\ket{010}-\ket{011})\\
    &\;\;\;\;\;\;\;\,+\gamma(\ket{100}+\ket{101})+\gamma(\ket{110}-\ket{111})\bigr).
\end{aligned}
\end{equation}
Instead of physically measuring in $X$-basis, a Hadamard gate $H$ is applied before its measurement in the computational $Z$-basis, where $H\ket{0}=\ket{+}=\frac{1}{\sqrt{2}}(\ket{0}+\ket{1})$, $H\ket{1}=\ket{-}=\frac{1}{\sqrt{2}}(\ket{0}-\ket{1})$. For consistency with the rest of the work, we also change the basis before measurements in this proof. Therefore, the state before measurement becomes
\begin{equation}
\begin{aligned}
    H_{1}\ket{\Phi}=&\frac{1}{2}\ket{0}_{1}\otimes((\alpha+\beta)\ket{00}+(\alpha-\beta)\ket{01}\\
    &\;\;\;\;\;\;\;\;\;\;\;+(\gamma+\delta)\ket{10}+(\gamma-\delta)\ket{11})_{0,2}\\
    &\!\!\!\!\!\!\!+\frac{1}{2}\ket{1}_{1}\otimes((\alpha-\beta)\ket{00}+(\alpha+\beta)\ket{01}\\
    &\;\;\;\;\;\;\;\;\;\;\;+(\gamma-\delta)\ket{10}+(\gamma+\delta)\ket{11})_{0,2},
\end{aligned}
\end{equation}
which reduces to
\begin{equation}
\begin{aligned}
    H_{1}\ket{\Phi}&=\frac{1}{2}\ket{0}_{1}\otimes H_{2}(\alpha\ket{00}+\beta\ket{01}+\gamma\ket{10}+\delta\ket{11})\\
    &\,+\frac{1}{2}\ket{1}_{1}\otimes X_{2}H_{2}(\alpha\ket{00}+\beta\ket{01}+\gamma\ket{10}+\delta\ket{11}).
\end{aligned}
\end{equation}
When the measurement of qubit $1$ in Pauli-$X$ basis is carried out, the superposition of state $\ket{0}$ and $\ket{1}$ on qubit 1 collapses to $\ket{s}$,
\begin{equation}\label{eq: measurement operator}
    M_{1,s}H_{1}\ket{\Phi}=\ket{s}_{1}\otimes X_{2}^{s}H_{2}\ket{\phi}_{0,2}.
\end{equation}
where the measurement operator $M_{i,x}$ projects the state on qubit $i$ of a multi-qubit state $\ket{a}$ in the $\ket{x}$ direction and $x\in\{0,1\}$, which is an eigenstate of the Pauli-$Z$ operator. When acted on an arbitrary state $\ket{a}$, it is denoted as $M_{i,x}\ket{a}=\frac{\ket{x}_{i}\bra{x}_{i}}{\|\braket{x}{a}\|^{2}}\ket{a}$. The state $\ket{\phi}_{0,2}$ now carries local transformation $X^{s}_{2}H_{2}$ \cite{Surface_codes}. Therefore, we have shown that for arbitrary two-qubit state $\ket{\phi}$, the claim as stated in \Cref{thm:teleportation across one qubit} holds. Or equivalently, 
\begin{equation}\label{eq:one hop teleportation}
    M_{1,s}H_{1}\underbrace{(CZ_{1,2}\ket{\phi}_{0,1}\otimes\ket{+}_{2})}_{\ket{\Phi}}=\ket{s}_1\otimes \underbrace{X_{2}^{s}H_{2}\ket{\phi}_{0,2}}_{\ket{\psi}^{\text{(Transformed)}}_{0,2}}.
\end{equation}
\end{proof}

\begin{customthm}{A2}\label{thm:teleportation across multiple qubit}
    \textit{Let $\ket{\Phi}=\left(\prod\limits_{i=1}^{n-2}{CZ_{i,i+1}}\right)\bigotimes\limits_{j=1}^{n-2}\ket{+}_{j+1}\otimes\ket{\phi}_{0,1}$ be a $n$-qubit quantum state with $\ket{\phi}_{0,1}$ as an arbitrary two-qubit quantum state rest on qubit 0 and 1 entangled with other qubits in the $X$-basis state $\ket{+}$ via $CZ$ gate in a path, where $CZ$ is the two-qubit entangling Controlled-$Z$ gate.}

    \textit{If all intermediate qubits from 1 to $n-2$ are measured in the $X$-basis with outcome $\vec{s}=(s_{1},s_{2},\cdots,s_{n-2})^{T}$, the state $\ket{\phi}_{0,1}$ will be teleported to qubit 0 and $n-1$ up to the local transformation that depends on the outcome $\vec{s}$, yielding }
    \begin{equation}
        \ket{\Phi}^{\text{Teleported}}=\left(\prod\limits_{i=1}^{n-2}X_{n-1}^{s_{i}}H_{n-1}\right)\ket{\phi}_{0,n-1}\otimes\ket{\vec{s}},
    \end{equation}
    \textit{where $\vec{s}\in\{0,1\}^{\otimes n-2}$.}
\end{customthm}
\begin{proof}[Proof.]
    From \Cref{thm:teleportation across one qubit}, we modify \Cref{eq:one hop teleportation} by changing labels and adding a new qubit to the system, such that the Hilbert space concerning the new equation is $\mathscr{H}_{N-3}\otimes\mathscr{H}_{0,N-2,N-1}$,
\begin{equation}\label{eq: initial case}
\begin{aligned}
    M_{n-2,s_{n-2}}H_{n-2}CZ_{n-2,n-1}\ket{s_{n-3}}\otimes \ket{\phi}_{0,n-2}\otimes\ket{+}_{n-1}\\
    =\ket{s_{n-3}}\otimes\ket{s_{n-2}}\otimes X_{n-1}^{s_{n-2}}H_{n-1}\ket{\phi}_{0,n-1},
\end{aligned}
\end{equation}
The newly added qubit $n-3$ is in the state $\ket{s_{n-3}}$ with $s_{n-3}\in \{0,1\}$, which is not coupled to any other qubits. Without loss of generality, we let
\begin{equation}\label{eq: substitution to RH}
    \ket{s_{n-3}}\otimes\ket{\phi}_{0,n-2}=\ket{s_{n-3}}\otimes X_{n-2}^{s_{n-3}}H_{n-2}\ket{\phi'}_{0,n-2},
\end{equation}
where $\ket{\phi'}$ is some two-qubit state can directly transform to $\ket{\phi}$ depends on the value of the newly added qubit, $s_{n-3}$. After change the labels, we substitute \Cref{eq:one hop teleportation} into RHS of \Cref{eq: substitution to RH}, yielding
\begin{equation}\label{eq: recursion}
\begin{aligned}
    \ket{s_{n-3}}\otimes\ket{\phi}_{0,n-2}=M_{n-3,s_{n-3}}H_{n-3}(CZ_{n-3,n-2}\ket{\phi'}_{0,n-3}\\
    \otimes\ket{+}_{n-2}),
\end{aligned}
\end{equation}
which is substituted back to LHS of \Cref{eq: initial case}. It is now clear that we can interpret $\ket{s_{n-3}}$ as the state collapsed after measuring in Pauli $X$-basis. Also, by applying the same trick that changes the label of qubit $n-2$ to $n-1$, \Cref{eq: substitution to RH} becomes
\begin{equation}\label{eq: substitution to RH_1}
    \ket{s_{n-3}}\otimes\ket{\phi}_{0,n-1}=\ket{s_{n-3}}\otimes X_{n-1}^{s_{n-3}}H_{n-1}\ket{\phi'}_{0,n-1},\vspace{10pt}
\end{equation}
which is substituted back to RHS of \Cref{eq: initial case}. By looping over the process of adding new qubits with the state $\ket{s_{n-4}},\ket{s_{n-5}},\cdots,\ket{s_{1}}$ to both sides of \Cref{eq: initial case} and recursively substituting \Cref{eq: recursion} to the LHS of \Cref{eq: initial case} and \Cref{eq: substitution to RH_1} to RHS of \Cref{eq: initial case}, we obtain
\begin{equation}
\begin{aligned}
    \ket{s_{1}\cdots s_{n-2}}\otimes\left(\prod\limits_{i=1}^{n-2}{X_{n-1}^{s_{i}}H_{n-1}}\right)\ket{\phi^{(n-2)}}_{0,n-1}&=
    \left(\prod\limits_{i=1}^{n-2}{M_{i,s_{i}}H_{i}CZ_{i,i+1}}\right)\bigotimes\limits_{j=2}^{n-1}\ket{+}_{j}\otimes\ket{\phi^{(n-2)}}_{0,1}
    \\
    &=\left(\prod\limits_{i=1}^{n-2}{M_{i,s_{i}}H_{i}}\right)\left(\prod\limits_{j=1}^{n-2}{CZ_{j,j+1}}\right)\bigotimes\limits_{k=2}^{n-1}{\ket{+}_{k}}\otimes\ket{\phi^{(n-2)}}_{0,1},
\end{aligned}
\end{equation}
where the Controlled-$Z$ operators are grouped to the right since they all commute with the Hadamard and measurement projection gates acting independently on different qubits. Similarly to \Cref{eq: substitution to RH_1}, we let $\ket{\phi}$ be a transformation of some state $\ket{\phi^{(n-2)}}$ after $n-2$ iterations,
\vspace{-1pt}
\begin{equation}
    T:S \rightarrow \mathscr{H'}, \ket{\phi}=(X^{s_{n-2}}H)\cdots(X^{s_{1}}H)\ket{\phi^{(n-2)}},
\end{equation}
where $S \subset\mathscr{H'}$ is some subset of the Hilbert space $\mathscr{H'}$, and recall that $\mathscr{H'}$ is spanned by all possible two-qubit state satisfying $\braket{\phi}=1$. Suppose $S$ takes the maximal domain $S=\mathscr{H'}$, then the image of $T$  is in some $D\subset \mathscr{H'}$. However, we know $D=\mathscr{H'}$ is strictly the full Hilbert space since $\ket{\phi}$ is set to be arbitrary. Thus, by the unitarity of $T$, the image of $T^{-1}$ must be equal to $\mathscr{H'}$. Or equivalently, the choice of $\ket{\phi^{(n-2)}}$ is also arbitrary. Therefore, we have proved \Cref{thm:teleportation across multiple qubit}.
\end{proof}

\section{Optimal paths search algorithm}\label{appendix B}
\vspace{-5pt}
\renewcommand{\thealgorithm}{B\arabic{algorithm}}
\begin{algorithm}[H]
    \caption{Optimal paths search algorithm}
    \label{alg: optimal paths}
    \textbf{Input}: A weighted graph $G$ with vertex set $V(G)$ and edge set $E(G)$, such that the edges are the available connections between the qubits weighted using one of the protocols discussed in \Cref{sec: Teleportation path optimisation}.\\
    \textbf{Output}: The best $m$ paths with $n$ qubits.
    \begin{algorithmic}
        \State Let $P=\{P_n=(v_1,\cdots,v_n)|P_n\subset G, d_{G}(v_1,v_n)=n-1\}$ be the set of paths that are all subgraphs of $G$ with $n$ vertices/qubits.
        \State Let $S = \emptyset$
        \For{$P_n \in P$}:
            \State $k = 1$
            \For{$e_{ij}\in E(P_n)$}:
                \State $k \leftarrow w(e_{ij})k$
            \EndFor
            \State $S\leftarrow S\cup\{(k, P_n)\}$
        \EndFor
        \State Let $T=(S, \le)$ be the ordered set of $S$ such that $\forall (k_1, P_{n_1}), (k_2, P_{n_2})\in S$, $k_1\le k_2\iff (k_1, P_{n_1})\le (k_2, P_{n_2})$.
        \State Let $T'$ be the last $m$ elements of $T$, where $P_{n}^{\text{(opt)}}\in T'$ is the last element.
        \State Output $T'$
    \end{algorithmic}
\end{algorithm}

\newpage
\section{Teleported state fidelity}\label{appendix C}
\begin{figure*}[htbp]
    \subfloat[\label{fig:Dynamic Circuits Teleported Fidelity}Dynamic Circuits without QREM]{\:\hspace{-7pt}\includegraphics[width=0.5\columnwidth]{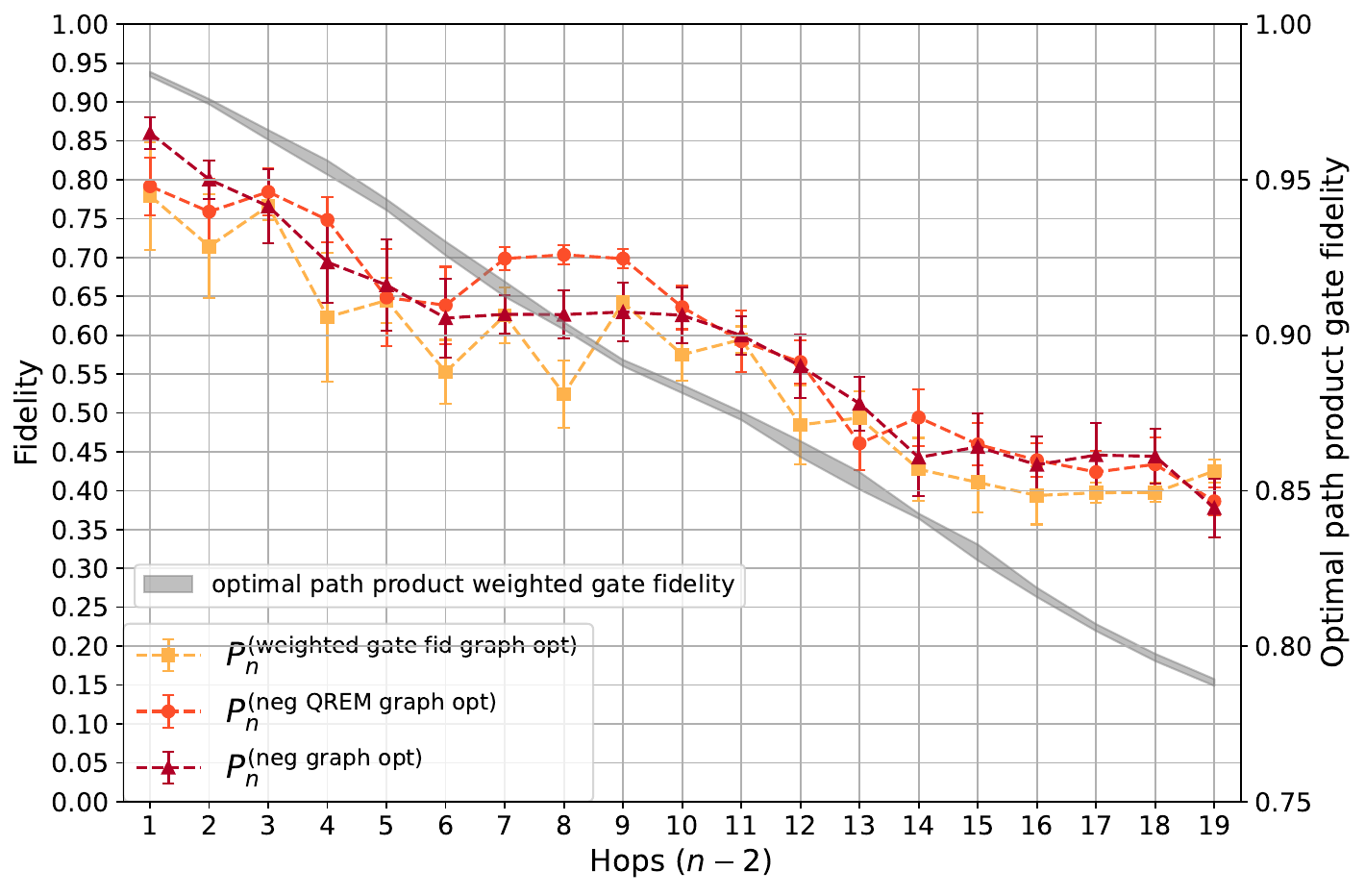}\:}
    \subfloat[\label{fig:Dynamic Circuits Teleported Fidelity QREM}Dynamic Circuits with QREM]{\:\includegraphics[width=0.5\columnwidth]{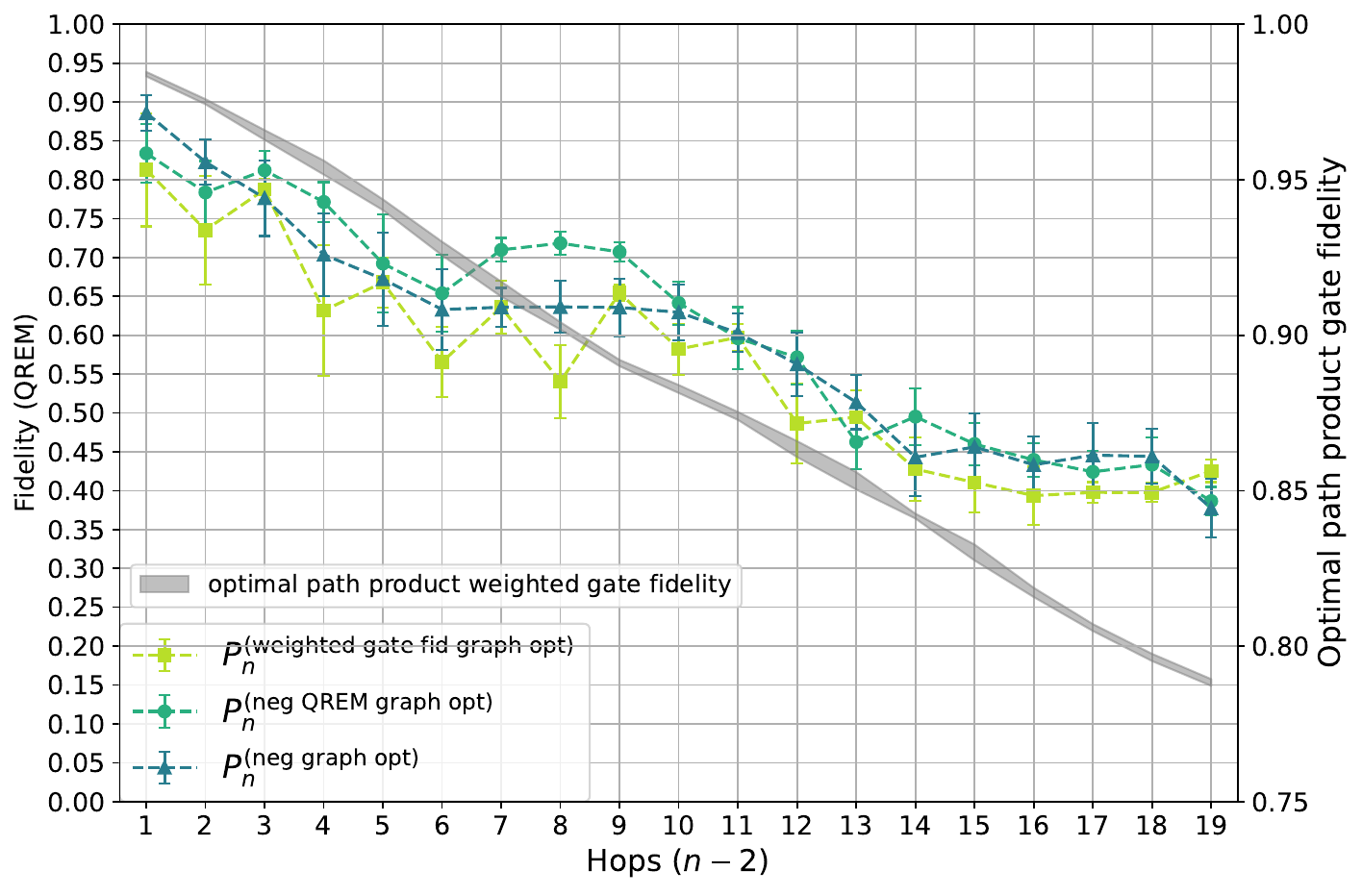}\:}\\
    \vspace{-10pt}
    \subfloat[\label{fig:Post Selected Teleported Fidelity}Post-Selection without QREM]{\hspace{-7pt}\:\includegraphics[width=0.5\columnwidth]{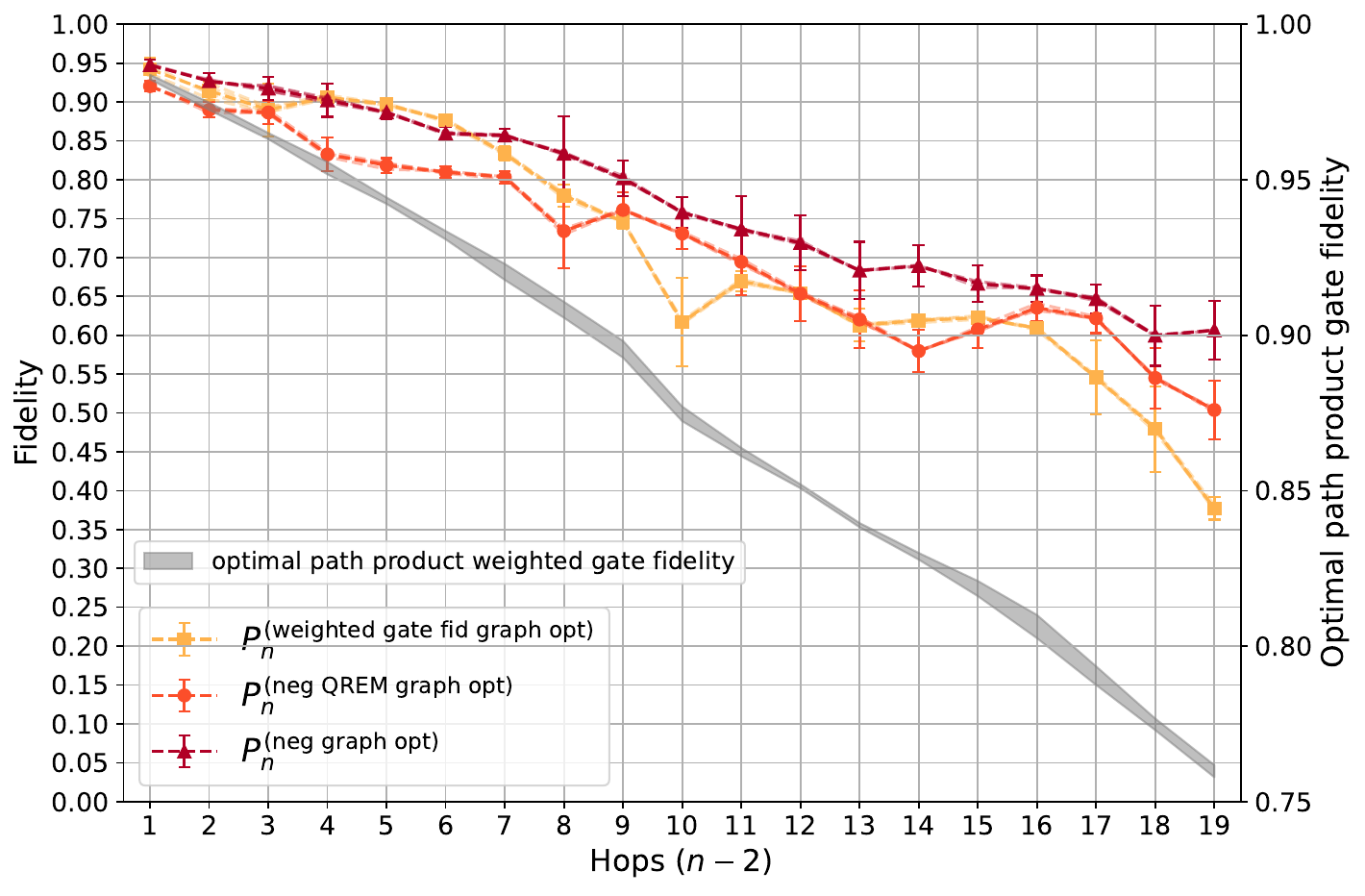}\:}
    \subfloat[\label{fig:Post Selected Teleported Fidelity QREM}Post-Selection with QREM]{\:\includegraphics[width=0.5\columnwidth]{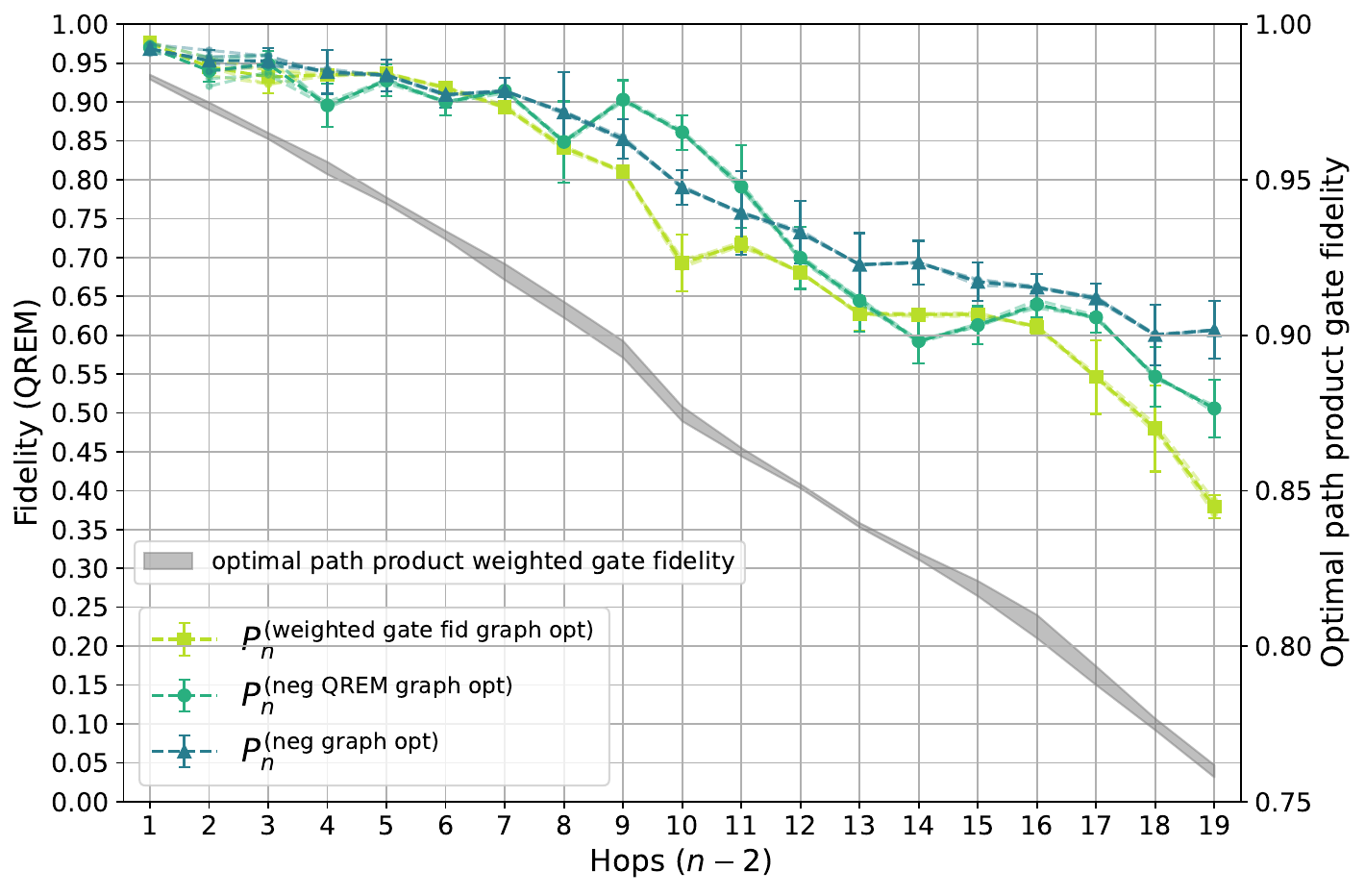}\:}
    \caption{\label{fig: fidelity plots detailed} Plots showing the results obtained for both dynamic circuit and post-selection approach as in \Cref{fig:dynamic circuit plots} and \Cref{fig:post processed plots} but evaluated in fidelity \textbf{(a)} The fidelity without QREM versus the number of qubits hopped in teleportation using dynamic circuits. \textbf{(b)} The fidelity with QREM versus the number of qubits hopped in teleportation using dynamic circuits. \textbf{(c)} The fidelity without QREM versus the number of qubits hopped in teleportation using post-selection. \textbf{(d)} The fidelity with QREM versus the number of qubits hopped in teleportation using post-selection.}
\end{figure*}

\newpage
\section{Path score comparisons using VF2 cost functions}\label{appendix D}

The VF2 algorithm finds its optimal paths by calculating and sorting the cost function that takes all gate errors and measurement errors after compilation into consideration, implemented using the Python package \texttt{mapomatic} \cite{VF2++, mapomatic}. In our case, the teleportation circuit (without QST) of $n$ qubits is first compiled into the device's native gate set for all possible paths $P_{n}'\subset G$. The optimal path(s) is then sorted based on the product of all involved gates and measurement errors corresponding to each of the compiled circuits. Therefore, the optimising condition of $P^{\text{(VF2 opt)}}_{n}$ becomes
\begin{equation}\label{eq:optimal paths conditions mapomatic}
    \mathcal{C}\left(O\left(P^{\text{(opt)}}_{n}\right)\right) = 1-\max_{P'_{n}\in G}\mathcal{F}\left(O\left(P'_{n}\right)\right)
\end{equation}
where $\mathcal{C}\left(O\left(P\right)\right)$ is the cost function score and 
\begin{equation}\label{eq: mapomatic cost function}
    \mathcal{F}\left(O(P)\right)=\prod\limits_{o_i\in O(P)}{(1-\varepsilon(o_i))}
\end{equation}
is the `net fidelity' for a given set of operations $O(P)$ transpiled on the path of qubits $P$. We choose the paths $P'_{n}=(v_1,\cdots,v_n)$ from path subgraphs of $G$ with length $n-1$ which also satisfies $d_{G}(v_1,v_n)=n-1$ (same as in \Cref{eq: optimal path}). $O(P_{n}') = \{o_i\}$ is the set of gates after compiling the teleportation circuit onto the path $P'_{n}$ and $\varepsilon(o_i)$ is the error rate corresponding to the operation $o_i$ on its specific qubit mapping. This algorithm is implemented through the Python package \texttt{mapomatic} \cite{VF2++, mapomatic}, and we label the corresponding optimal paths as $P^{\text{(VF2 opt)}}_{n}$. 

Similarly, we also design another protocol with optimal paths labelled as $P^{\text{(neg VF2 opt)}}_{n}$ that is exactly the same as above except that $\varepsilon(o_i)$ when $o_i$ is the two-qubit gate is replaced by the negativity (with QREM) between the two qubits that this gate is acting on.

\begin{figure*}[htbp]
    \subfloat[\label{fig:Scores mapomatic}VF2 cost function scores for paths on dynamic circuits]{\hspace{-7pt}\:\includegraphics[width=0.5\columnwidth]{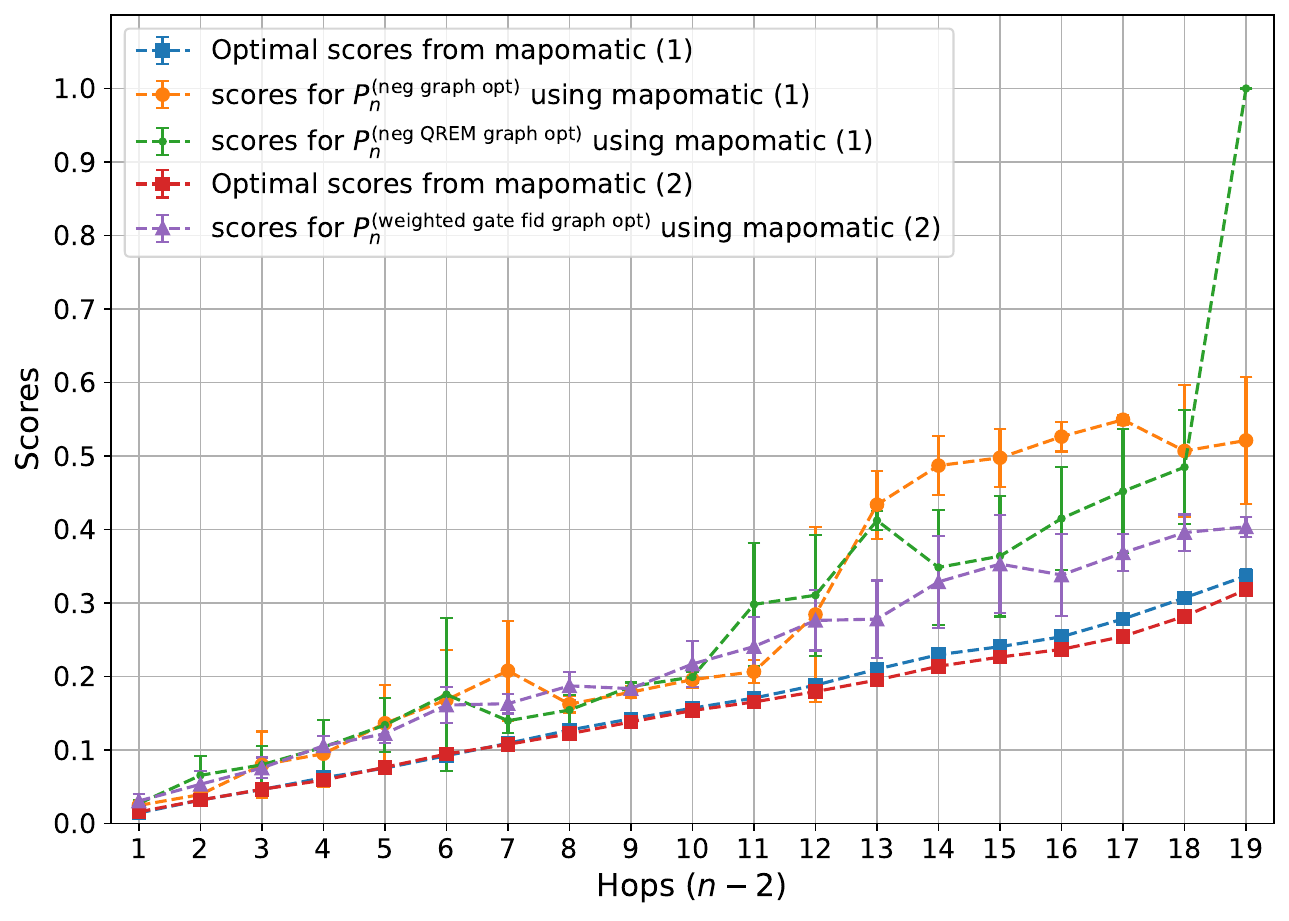}\:}
    \subfloat[\label{fig:gate fid opt paths scores mapomatic}VF2 cost function scores for weighted gate fidelity optimal paths]{\:\includegraphics[width=0.5\columnwidth]{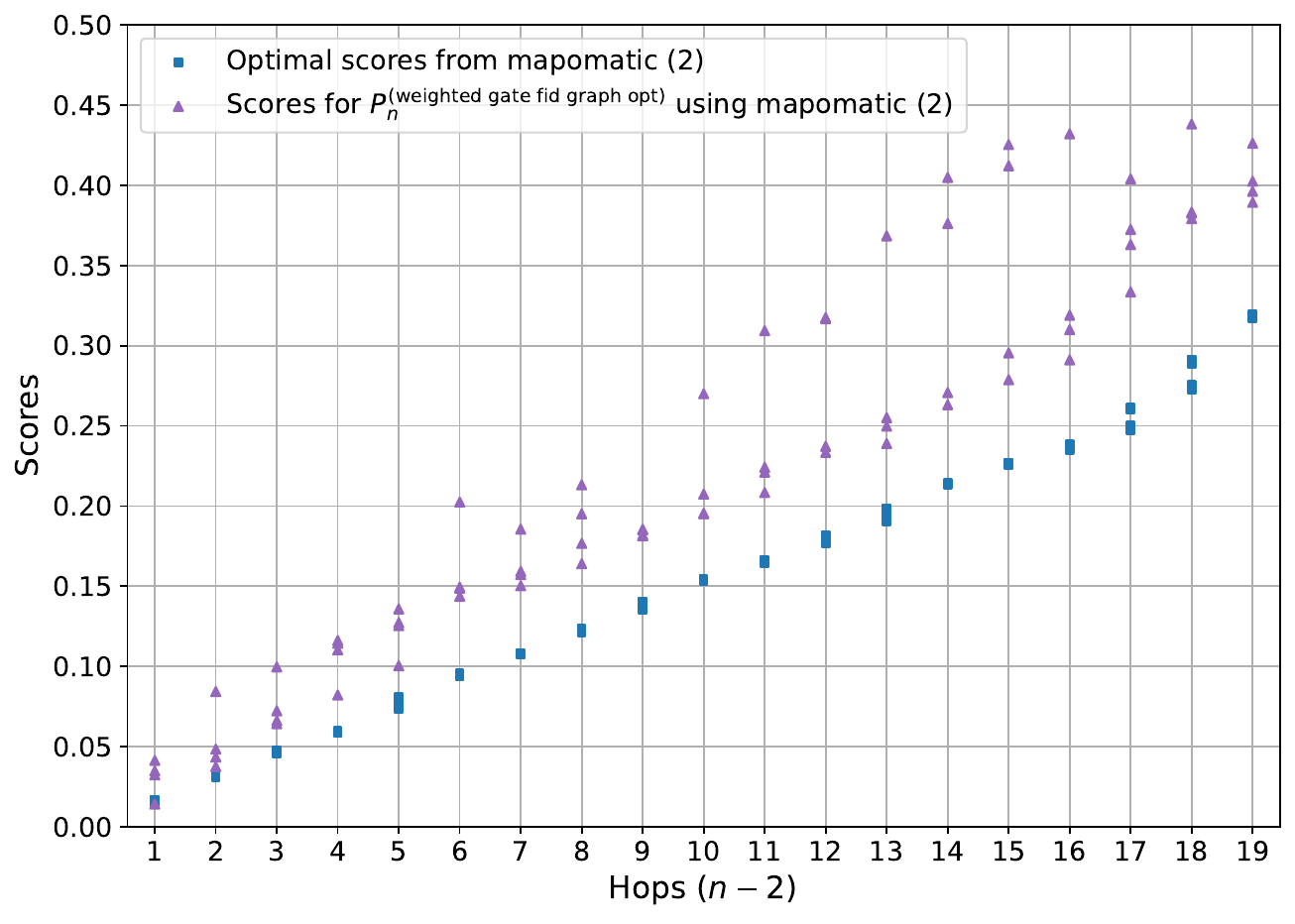}\:}
    \caption{\label{fig: Scores} Plots showing the scores evaluated from the VF2 cost functions. According to \Cref{eq:optimal paths conditions mapomatic}, a lower score is better. `Optimal scores from mapomatic' labels the average score of the best four paths found using the algorithm as in \Cref{eq:optimal paths conditions mapomatic}, while the others label the average scores of the four paths $P_n^{\text{(neg graph opt)}}$, $P_n^{\text{(neg QREM graph opt)}}$, $P_n^{\text{(weighted gate fid graph opt)}}$ evaluated using the cost function \Cref{eq: mapomatic cost function}. The numbers in parentheses, (1) and (2), denote the calibration data retrieved at the same time. \textbf{(a)} A comparison between each of the pathfinding approaches. An outlier is observed in scores for $P_n^{\text{(neg QREM graph opt)}}$, which could be due to the two-qubit gate error rate falsely labelled as 1 yet behaving properly when benchmarked for negativity. \textbf{(b)} More detailed plot showing the scores of all four paths for 'Optimal scores from mapomatic (2)' and $P_n^{\text{(weighted gate fid graph opt)}}$. 
    }
\end{figure*}
\end{document}